\documentclass{aa}

\usepackage{graphicx}
\usepackage{txfonts}
\usepackage{amsmath}
\usepackage{amssymb}
\usepackage{graphicx}
\usepackage{textcomp}
\usepackage{gensymb}
\usepackage{listings}

\usepackage{subcaption}
\usepackage{multirow}
\usepackage[normalem]{ulem}
\usepackage{lscape}
\usepackage{longtable}
\usepackage{booktabs}
\usepackage{pdfpages}
\usepackage[flushleft]{threeparttable}
\usepackage[hidelinks]{hyperref}

\bibpunct{(}{)}{;}{a}{}{,} 

\begin{document}

\title{HI observations of the MATLAS dwarf and ultra-diffuse galaxies\thanks{Table \ref{tab:properties} is only available in electronic form at the CDS via anonymous ftp to cdsarc.u-strasbg.fr (130.79.128.5) or via \url{http://cdsweb.u-strasbg.fr/cgi-bin/qcat?J/A+A/}.}}
\author{M\'elina Poulain\inst{1}, Francine R. Marleau\inst{1}, Rebecca Habas\inst{2}, Pierre-Alain Duc\inst{2}, Rub{\'e}n S{\'a}nchez-Janssen$^{3}$, Patrick R. Durrell$^{4}$, Sanjaya Paudel$^{5}$, Oliver M{\"u}ller$^{2}$, Sungsoon Lim$^{6}$, Michal B{\'i}lek$^{7}$, J{\'e}r{\'e}my Fensch$^{8}$}
\titlerunning{HI observations of the MATLAS dwarfs}
\authorrunning{M. Poulain et al.}

\institute{Institute f{\"u}r  Astro- und Teilchenphysik, Universit{\"a}t Innsbruck, Technikerstra{\ss}e 25/8, Innsbruck, A-6020, Austria
\\e-mail: melina.poulain45@gmail.com, Melina.Poulain@student.uibk.ac.at
\and Observatoire Astronomique, Universit{\'e} de Strasbourg, CNRS, 11, rue de l'Universit{\'e}. F-67000 Strasbourg, France
\and UK Astronomy Technology Centre, Royal Observatory Edinburgh, Blackford Hill, Edinburgh EH9 3HJ, UK
\and Dept. of Physics, Astronomy, Geology, and Environmental Sciences, Youngstown State University, Youngstown, OH 44555 USA
\and Department of Astronomy and Center for Galaxy Evolution Research, Yonsei University, Seoul 03722
\and Department of Astronomy, Yonsei University, 50 Yonsei-ro Seodaemun-gu, Seoul, 03722, Republic of Korea
\and Nicolaus Copernicus Astronomical Center, Polish Academy of Sciences, Bartycka 18, 00-716 Warsaw, Poland
\and Univ. Lyon, ENS de Lyon, Univ. Lyon 1, CNRS, Centre de Recherche Astrophysique de Lyon, UMR5574, 69007 Lyon, France}

\date{}

\abstract
{
The presence of HI gas in galaxies is inextricably linked to their morphology and evolution. This paper aims to understand the HI content of the already identified 2210 dwarfs located in the low-to-moderate density environments of the MATLAS deep imaging survey. We combine the HI observations from the ATLAS$^{3D}$ survey, with the extragalactic HI sources from the ALFALFA survey, to extract the HI line width, velocity and mass of the MATLAS dwarfs. From the 1773 dwarfs in our sample with available HI observations, 8\% (145) have an HI line detection. The majority of the dwarfs show irregular morphology, while 29\% (42) are ellipticals, the largest sample of HI-bearing dwarf ellipticals (dEs) to date. Of the HI dwarf sample, 2\% (3) are ultra-diffuse galaxies (UDGs), 12\% have a transition-type morphology, 5\% are tidal dwarf candidates, and 10\% appear to be disrupted objects. In our optically selected sample, 9.5\% of the dEs, 7\% of the UDGs and 10\% of the classical dwarfs are HI-bearing. The HI-bearing dwarfs have on average bluer colors than the dwarfs without detected HI. We find relations between the stellar and HI masses, gas fraction, color and absolute magnitude consistent with previous studies of dwarfs probing similar masses and environments. For 79\% of the dwarfs identified as satellites of massive early-type galaxies, we find that the HI mass increases with the projected distance to the host. Using the HI line width, we estimate dynamical masses and find that 5\% (7) of the dwarfs are dark matter deficient.
}


\maketitle

\section{Introduction}

In recent years, our understanding of the processes that govern galaxy growth and evolution have expanded to include both internal and external drivers, the so-called "nature" versus "nurture" problem.
One piece of evidence supporting the role of the environment on galaxy morphology is the morphology-density relation \citep{Dressler1980}, where the early-type galaxies (ETGs), the red passive ellipticals and lenticulars, are more likely found in high density cluster environments, while the late-type galaxies (LTGs), comprising the blue star-forming spirals and irregulars, inhabit less crowded environments. Tidal features, such as tails, shells and streams, caused by galaxy interactions (e.g., \citealt{Malin1980,Schweizer1982,Tal2009,Janowiecki2010,Hood2018,Mueller2019}) offer additional evidence for environmental processes acting on galaxies. Studies of the quenching of the star-forming galaxies have shown that the star-forming activity can be regulated by both environmental (e.g., tidal and ram-pressure stripping, harassment, mergers) and internal processes (e.g., AGN and supernovae feedback, gravitational quenching). The formation of the central region, or bulge, of galaxies also depends on both external and internal effects, where the classical bulges are the outcome of mergers \citep{Aguerri2001} and the pseudo-bulges are produced by the gas internally driven by bars \citep{Kormendy2004}. 

While the "nature" versus "nurture" problem has long been studied in massive galaxies (stellar mass M$_* \gtrsim 10^9$M$_{\odot}$), the full impact of the environment has not been explored for the less massive galaxies, also known as dwarf galaxies.
Until recently, most studies of dwarfs focused on nearby clusters, groups in the Local Volume (LV) and the Local Group (LG), leading to a lack of observations in low density environments. We know that the local environment can be linked to the morphological type of the dwarfs which can be divided into two main groups: the dwarf irregulars (dIs), with irregular isophotes, high gas fraction and ongoing star formation, and the dwarf ellipticals/spheroidals (dEs/dSphs) characterized by a lack of gas and star formation and their elliptical isophotes. Similar to more massive galaxies, we observe a morphology-density relation for dwarf galaxies, where dIs are located, on average, in less dense local environments than dEs \citep{Ferguson1989, Mcconachie2012, Skillman2003, Cote2009, Habas2020}. Studies of the morphology of dwarfs located in low-to-moderate density environments also suggest that the morphological type of galaxy satellites is linked to that of their massive host (either an ETG or LTG) and to the projected distance to this host, with dEs being more numerous around ETGs and located closer to their host than star-forming satellites \citep{Geha2012,Ann2017,Habas2020}.
Moreover, a number of dwarfs also show signs of galaxy interactions, and a catalog of dwarf galaxies exhibiting tidal features in the Local Universe is presented in \citet{Paudel2018}. These interacting dwarfs appear to favor low density environments. On the other hand, simulations of isolated dwarfs have demonstrated the role of internal processes in shaping dwarf morphologies. These hydrodynamical simulations, based on supernova feedback and ultraviolet radiation, produce dEs whose morphological properties are similar to those of LG dwarfs \citep{Valcke2008, Revaz2009, Revaz2018}.

Studies of the neutral gas (HI) content of massive galaxies provide clues about the effects of the external and internal processes on galaxy morphology. The impact of the environment can be observed from the fact that galaxies of similar stellar masses have less HI in clusters and groups than in the field \citep{Davies1973,Giovanelli1985,Verdes2001,Hess2013,Denes2014}. On the other hand, the relation between the gas fraction and the stellar mass is an example of the impact of internal processes in galaxies, where the HI-to-stellar mass ratio decreases towards galaxies of higher stellar masses \citep{Catinella2018}.

In HI studies of dwarfs, we also observe the effects of both internal and external processes. The relation between the HI-to-stellar mass ratio and the stellar mass extends to low mass galaxies \citep{Huang2012}. In the LG, the HI mass of M31 and the Milky Way satellites increases with the distance to the host and correlates with the morphological type of the dwarfs, the dEs having less HI gas and being located closer to the host as compared to the dIs \citep{Grebel2003,Grcevich2009}. Several studies of HI-bearing dwarfs (e.g., \citealt{Dellenbusch2008,Koleva2013}) also revealed the existence of an intermediate morphology, the transition-type dwarfs (TTDs), typically showing a low HI mass-to-light ratio, an elliptical shape with internal star-forming regions and an external area containing an old stellar population.

In the last few years, the role of nature versus nurture in shaping the so-called ultra diffuse galaxies (UDGs; defined as galaxies with very low central surface brightnesses $\mu_{g,0}$ $> 24$ mag/arcsec$^2$ and large effective radii $R_e > $ 1.5 kpc, \citealt{vanDokkum2015}), has been of particular interest. Some groups have argued that internal processes are dominant, e.g., AGN, supernova or stellar feedback \citep{DiCintio2017,Papastergis2017}. While others suggest that they are the result of an environmental effect such as tidal disruption \citep{Mihos2015,Merritt2016,Bennet2018} or collisions \citep{Baushev2018}. 
Studies of globular clusters and morphology of UDGs indicate multiple origins of UDGs even in one cluster \citep{Toloba2018,Lim2018,Lim2020}, and simulations support the idea of the existence of several populations of UDGs \citep{Sales2020}. The absence of environmental preferences for HI-bearing UDGs observed by \citet{Janowiecki2019} favors a formation of UDGs based on internal processes. Coupled to the study of HI-bearing UDGs in low density environments of \citet{Leisman2017}, their results suggest that gas-rich UDGs could undergo gas-stripping and evolve into the observed quiescent cluster UDGs during their infall.

The work by \citet{Habas2020} on the deep optical images of the Mass Assembly of early-Type GaLAxies with their fine Structures (MATLAS) survey unveiled a sample of 2210 dwarf galaxies in low-to-moderate density environments with a large majority of dEs ($\sim$73\%). Therefore, an advantage of an HI study of such sample is that we are not biased by the morphological type of the HI-bearing dwarfs, unlike some works that favor the observations of a dI population or target some peculiar dEs. Furthermore, since this sample is not HI selected, we are able to quantify the number of dwarfs of different types and UDGs with HI detection along with their HI-derived properties. Photometric and structural properties were extracted for 1589 dwarfs with the use of Sérsic modelling \citep{Poulain2021}, and a subsample of 59 UDGs was then defined based on these properties \citep{Marleau2021}. Recent work on the MATLAS dwarfs found evidence, for example, for a morphology-density relation \citep{Habas2020}, no statistically significant differences between the structural properties of dwarfs in different environments \citep{Poulain2021}, and similarities between the properties of UDGs and classical dwarfs, suggesting a similar formation path for both populations \citep{Marleau2021}.
In this paper, we investigate the HI content of the MATLAS dwarfs, in addition to their optical study. We compare the morphology, local environment and scaling relation of the MATLAS HI-bearing dwarfs to existing HI dwarf samples.

This paper is organized as follows. In Section 2, we present the MATLAS dwarfs and UDGs samples. In Section 3, we describe the HI observations and the resulting catalogue of MATLAS HI dwarfs and UDGs. The HI fluxes and estimated masses as well as the optical properties of the HI sample are presented in Section 4. We discuss the results in Section 5 and our conclusions in Section 6.

\section{The MATLAS dwarf and UDG samples}
\label{section:MATLAS_sample}

The dwarf and UDG samples were created based on the optical images of the MATLAS survey \citep{Duc2015}.
This survey is a facet of the ATLAS$^{3D}$ project \citep{Cappellari2011}, which aims to characterize the morphology and the kinematics of ETGs in the context of galaxy formation and evolution, and whose observational part is composed of multi-wavelength observations (in radio, millimeter and optical) of a complete sample of 260 ETGs with distances $\lesssim$ 45 Mpc, declinations $|\delta - 29^\circ | < 35^\circ $, galactic latitudes $|b| > 15^\circ$, and K-band absolute magnitudes $M_K < -21.5$.
The optical component of ATLAS$^{3D}$ was undertaken by the MATLAS survey and the NGVS (Next Generation Virgo Survey, \citealt{Ferrarese2012}) with MegaCam on the 3.6~meter Canada-France-Hawaii Telescope (CFHT). The 58 ETGs located in the Virgo cluster were observed for NGVS between 2009 and 2014, while 150 $1^\circ \times 1^\circ$ fields containing 180 ETGs and 59 LTGs located outside Virgo, in low to moderate density environments, were targeted for MATLAS between 2012 and 2015. The MATLAS survey data were observed in the g,r,i bands for 150, 148 and 78 fields respectively as well as in the u band for the 12 fields with a distance below 20 Mpc. 

With surface brightnesses ranging down to $28.5 - 29$ mag/arcsec$^2$ in the g-band, and an average image quality of 0.96\arcsec, 0.84\arcsec, 0.65\arcsec and 1.03\arcsec$\ $ in the g, r, i and u bands respectively, the MATLAS images are suitable for the search and identification of dwarf galaxies. Combining a visual inspection of the 150 g-band field images and semi-automated catalogs based on {\sc{Source Extractor}} \citep{Bertin1996} output parameters, \citet{Habas2020} defined a preliminary catalog of 25522 dwarf galaxy candidates. These candidates were then reviewed in a two-step process.  First, each candidate was visually inspected by at least three team members and likely dwarfs were flagged and morphologically classified. The resulting 3311 flagged galaxies were subsequently inspected a second time by a five member team in order to remove any remaining potential background galaxies and to confirm the assigned morphological classifications.   After removing any remaining duplicates, this lead to a clean sample of 2210 dwarf galaxies. Based on their absolute magnitude, knowing that the brightest dwarfs have M$_B \sim -18$ \citep{Kormendy1985}, the dwarf nature was confirmed for 99\% of the galaxies with an available distance measurement (13.5\% of the sample, including some of the HI distances presented here) and the one-sky positions and relative velocities suggest that 90\% of this subsample are satellites of the nearest massive galaxy (ETG or LTG) in the ATLAS$^{3D}$ parent sample. In a similar test, 80\% of the subsample of dwarfs with known distances were shown to be satellites of the targeted ETG in the images. Based on these statistics, the rest of the dwarfs in the sample without known distances were assumed to be associated with the central ETG in the field and the distance of the ETG was therefore assigned to the dwarfs. Although the sample appears robust, it is likely incomplete. Galaxies with diffuse haloes but extended light concentrations in the nuclear region, irregular galaxies, small objects, as well as galaxies at the bright end of our initial sample were difficult to classify and hence some dwarfs were potentially rejected. It is particularly important to note for this work that some dIs, potentially HI-bearing, may have been missed due to their possible confusion with background galaxies. The MATLAS dwarf sample completeness is described in \citet{Habas2020}. A discussion specifically related to the HI dwarfs is presented in Section \ref{section:HIprop}.

An analysis of the structural and photometric properties of the MATLAS dwarfs was performed using the software \textsc{galfit} \citep{Peng2010}. The galaxies were fit by a Sérsic profile \citep{Sersic1963} and the structural and photometric properties, such as the Sérsic index, effective radius, the apparent magnitude and the $(g-r)$ and $(g-i)$ colors, were derived for 1589 MATLAS dwarf galaxies, including 1437 dEs and 152 dIs \citep{Poulain2021}. The surface brightness of the modeled dwarfs was computed based on the equations of \citet{Graham2005}, using the total magnitude and effective radius. We note that the properties were derived for 87\% of the dEs and 27\% of the dIs; we could not get a reliable model for a large majority of the dIs due to their irregular shape and internal structures, and for some dEs due to a very bright center. 

A subsample of the 1589 modeled dwarfs, 59 in total, have effective radii $\geq$ 1.5 kpc and a central surface brightness $\geq$ 24 mag/arcsec$^2$ \citep{Marleau2021}. We will adopt these galaxies as our UDG subsample for the remainder of the paper. 

\section{HI observations}

\subsection{ATLAS$^{3D}$ HI survey}
The ATLAS$^{3D}$ HI survey \citep{Serra2012} consists of the radio observations of 166 nearby ETGs from the ATLAS$^{3D}$ project with a declination $\delta$ > 10\degree$\ $~by the Westerbork Synthesis Radio Telescope (WSRT). We focus our study on the subsample of 110 ETGs situated outside the Virgo cluster, in common with the MATLAS survey. Of these 110, 14 were part of previous surveys (\citealt{Morganti2006}, \citealt{Oosterloo2010}) that studied the kinematics of SAURON ETGs with HI line observations with the WSRT \citep{Zeeuw2002}. In both the ATLAS$^{3D}$ and SAURON surveys, the galaxies have been observed for 12h over a bandwidth of 20 MHz ($\sim$4000 km s$^{-1}$) sampled with 1024 channels with a field of view of about one square degree\footnote{Except for the field of view of NGC4150 which is about 2 square degrees.}. The telescope has a detection limit of M$_{HI} \sim 5.5\times10^6$ M$_{\odot}$ for a galaxy as far as the Virgo cluster ($\sim$ 16 Mpc) and a beam size of $\sim$ 1\arcmin. To obtain the HI cubes\footnote{Available at \url{http://www-astro.physics.ox.ac.uk/atlas3d/}.}, 
the data were reduced with a pipeline based on the \textsc{miriad} package \citep{Sault1995}, and the resulting HI cubes were constructed using a robust null weighting \citep{Briggs1995} and a velocity resolution of 16 km s$^{-1}$ after Hanning smoothing.
 
\subsection{ALFALFA survey}
The ALFALFA survey data was used to complement the the ATLAS$^{3D}$ data. It is a second generation blind survey specialized in extragalactic HI sources detection with a better sensitivity and resolution than previous blind surveys such as the HI Parkes All Sky Survey (HIPASS, \citealt{Staveley-Smith1996}) and the HI Jodrell All-Sky Survey (HIJASS, \citealt{Lang2003}). The observations were performed with the Arecibo telescope over an area of $\sim$7000 deg$^2$ split in two regions with a declination 0\degree\ < $\delta$ < 36\degree\ and two right ascension ranges 07$^h$30$^m$ < RA < 16$^h$30$^m$ and 22$^h$ < RA < 03$^h$, with a beam size $\sim$4\arcmin. As it was designed to study the low mass end of the HI mass function \citep{Giovanelli2005}, it can detect galaxies with an HI mass down to $10^6$ M$_{\odot}$. This survey has a velocity resolution of 10 km s$^{-1}$ after Hanning smoothing. For our work, we use the final release of the extragalactic HI sources catalog \citep{Haynes2018} which contains approximately 31,500 sources, most with a SNR > 6.5. HI sources with a lower SNR were kept if they were successfully matched with a galaxy of known redshift.\\ 

\subsection{HI dwarf and UDG samples}

Of the 2210 dwarf galaxies in the MATLAS survey, 1773 dwarfs fall within the footprints of either the ATLAS$^{3D}$ HI survey (1141) or the ALFALFA survey (1139). It should be noted, however, that we do not expect a detection for each of these dwarfs as the dEs (77\%) are expected to be gas poor. Additionally, given the assumed distances of the dwarfs and the mass detection limits of the telescopes, it is unlikely that we will detect HI masses below $10^7$ M$_{\odot}$, $10^6$ M$_{\odot}$ from Arecibo, WSRT, respectively. Nevertheless, we searched for HI-counterparts for all 1773 dwarfs within the footprints of the two surveys, using the cubes and contours of the ATLAS$^{3D}$ survey as well as the ALFALFA extragalactic HI source catalog. First, we searched for HI detections with separations smaller than the beam size of the telescope from the optical detections. Then, we visually inspected the matched sources using g-band images and, in the case of the ATLAS$^{3D}$ HI survey, the HI images to ensure that the dwarf is the HI source and that the HI is not centered on a nearby galaxy. We detected HI in 145 dwarfs, with 94 and 64 detections in ATLAS$^{3D}$ and ALFALFA, respectively, and a total of 13 galaxies detected in both surveys. We note that this sample includes 42 dEs, and that 18\% are classified as nucleated. Of the HI sample, three dwarfs are classified as UDGs based on their structural properties. The HI line profiles of the 94 detections in ATLAS$^{3D}$, and 145 color cutouts of the galaxies are displayed in Figure \ref{fig:HIspectra} and \ref{fig:colorcutouts} of Appendix \ref{AppendixA}, respectively. We highlighted the dwarfs detected in both surveys with a red name in the HI spectra.

\begin{figure*}
\centering
\includegraphics[scale=0.42]{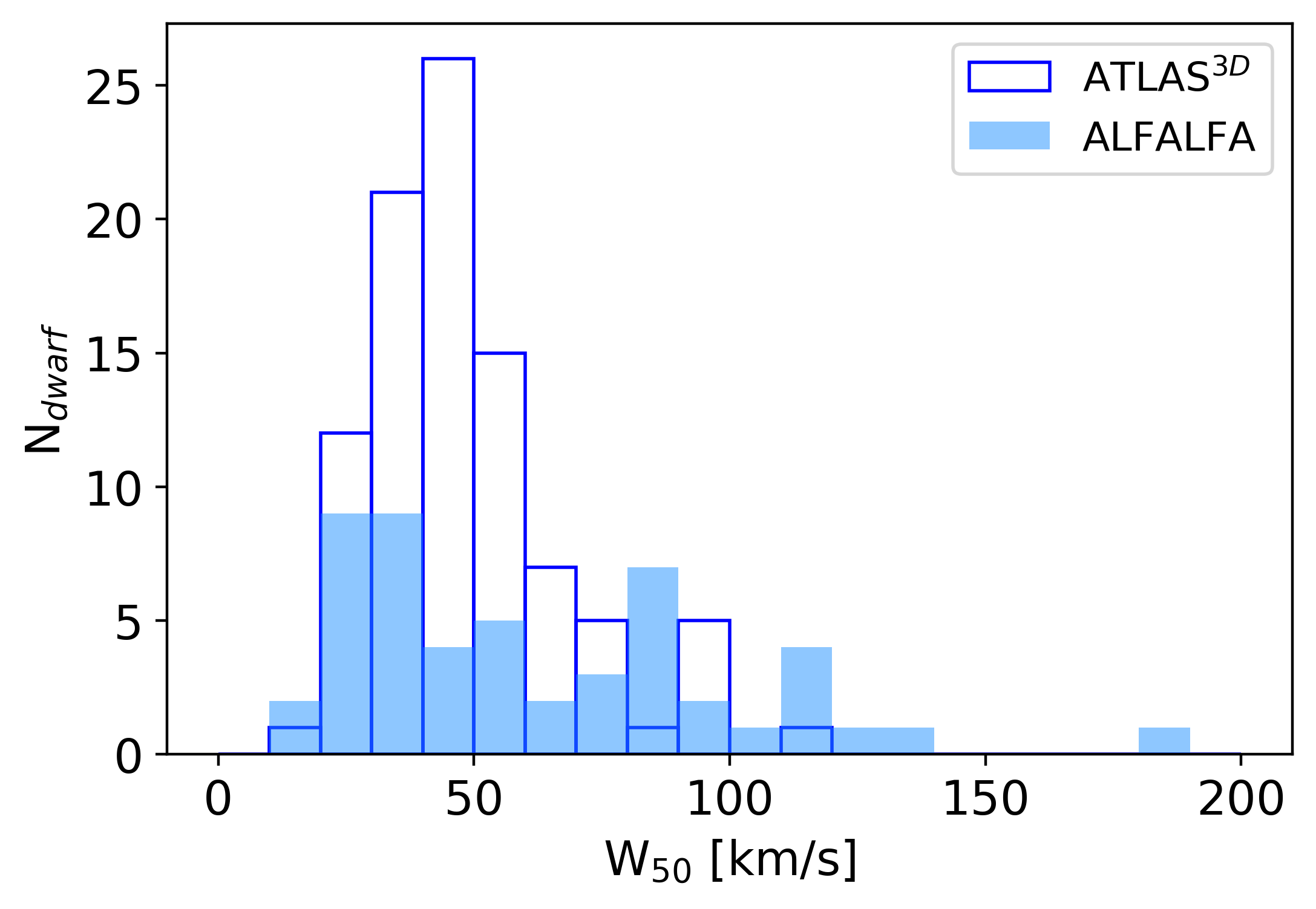}
\includegraphics[scale=0.42]{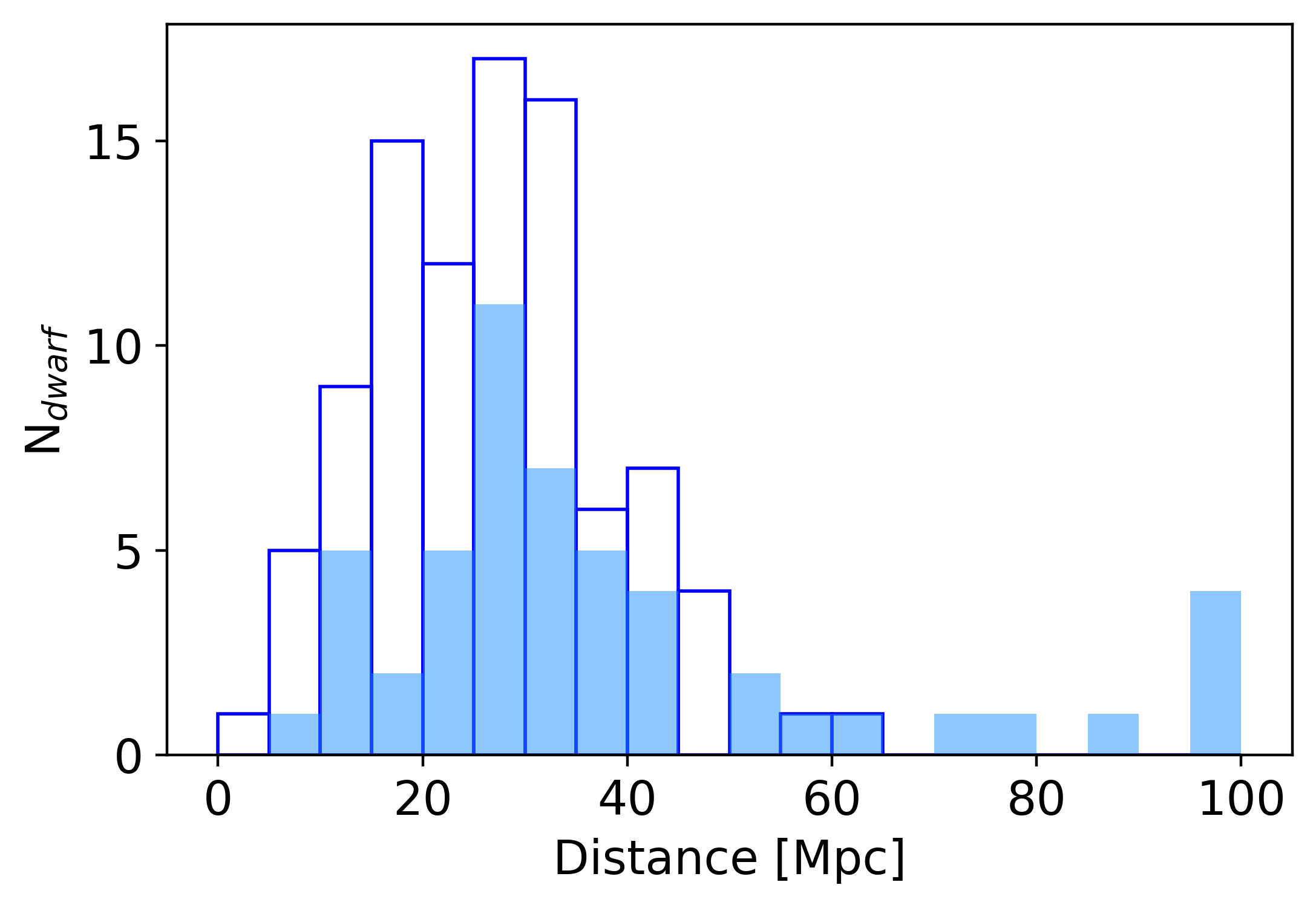}
\includegraphics[scale=0.42]{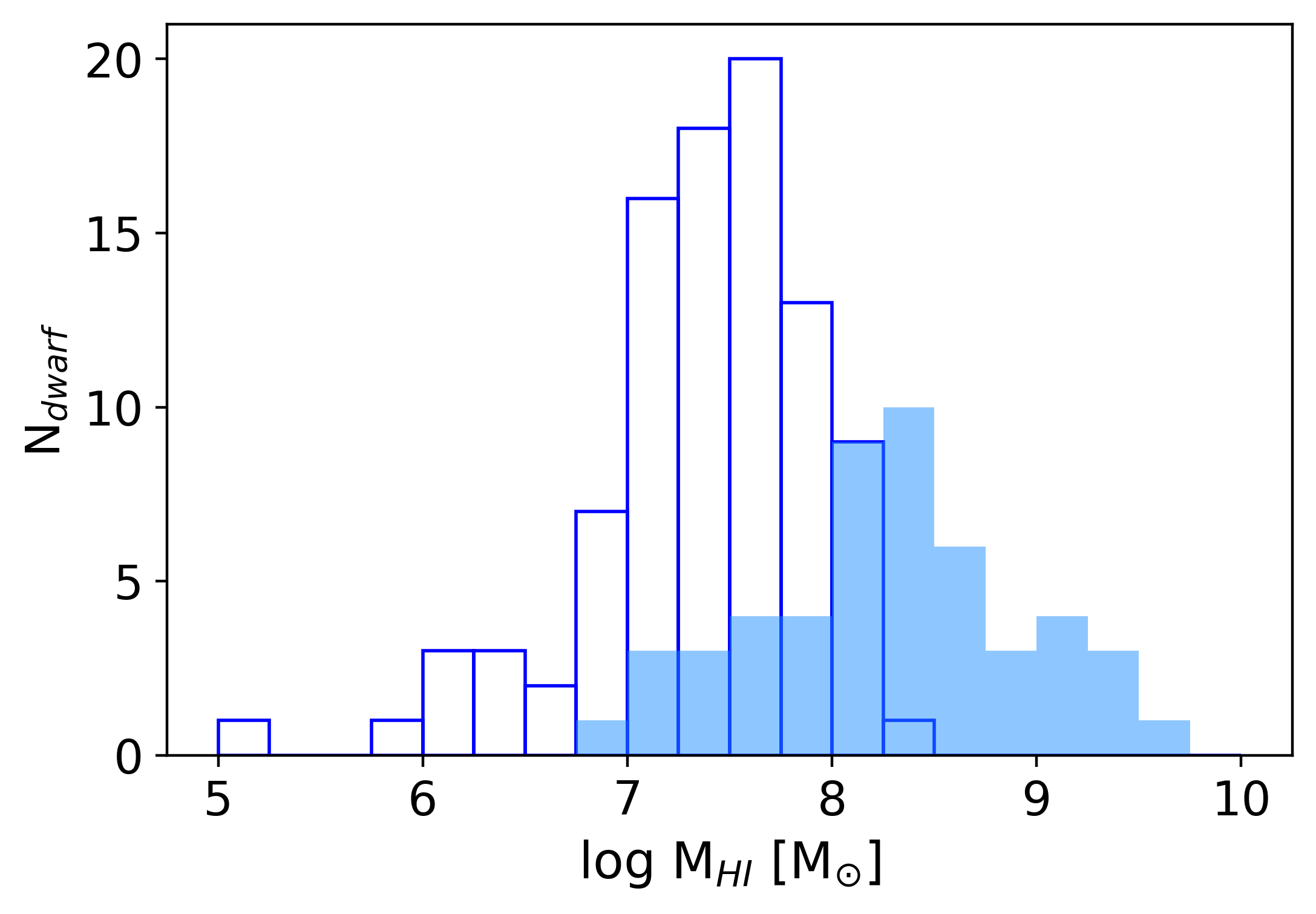}
\caption{Distributions of the HI properties of the MATLAS dwarfs sample. Empty bars: detection in the ATLAS$^{3D}$ HI survey. Open bars: detection in the ALFALFA survey.}
\label{fig:HIprop}
\end{figure*}

\section{HI and optical properties of the MATLAS dwarfs}
\subsection{HI properties}
\label{section:HIprop}

We present here the HI properties of the 145 HI-bearing dwarfs as well as the methods used to extract them from the HI spectra. The HI properties are available in Table \ref{tab:properties}.\\

\textbf{W$_{50}$}. Defined as the line width measured at 50\% of the maximum intensity. We measure W$_{50}$ in the range 19$-$186 km/s. One dwarf, MATLAS-447, has a W$_{50}$ value of 16 km/s, which is roughly the same as the velocity resolution of the ATLAS$^{3D}$ HI survey; we have kept the galaxy in the sample, but will not include it in any discussion of W$_{50}$ going forward.\\

\textbf{Velocity}. The velocity corresponds to the median point of the 21 cm line profile measured at 50\% of maximum intensity. The MATLAS HI dwarfs show a large range of velocities, from 200 to 7000 km/s. We estimate errors on the velocities in the range $4.1-12.4$ km/s.\\ 

\textbf{Distance}. We corrected all obtained heliocentric velocities for the infall to the Virgo cluster and to the Local Group following the method from \citet{Mould2000}. From these corrected velocities, we computed the distance of the galaxy using the formula $D=v/H_0$ with $H_0$ = 70 km s$^{-1}$ Mpc$^{-1}$ and estimated its uncertainty. To be consistent, we computed the distances using the same method for both ATLAS$^{3D}$ and ALFALFA extracted velocities, as the distance to the galaxies with a velocity below 6000 km/s were obtained using a different method in the ALFALFA catalog. The distances we compute are between 2 and 99 Mpc. These HI distances are used to calculate all the distance-dependent HI and optical properties of this paper.\\

\textbf{HI mass and flux}. We derived the HI mass from the standard formula log(M$_{HI}$/M$_{\odot}$)=$2.356\times10^5D^2F_{HI}$ \citep{Roberts1962}, with $D$ the derived distance in Mpc and $F_{HI}$ the integrated HI flux in Jy km.s$^{-1}$. The HI spectrum of each dwarf has been fit by a Gaussian and then integrated to get the flux value $F_{HI}$. The MATLAS dwarfs with HI detections have HI masses log(M$_{HI}$/M$_{\odot}) \lesssim 9.5$. Similar to the method from \citep{Haynes2018}, we compute the error on the HI mass based on the error on the HI flux, derived from the uncertainties of the Gaussian fit, and the error on the distance.\\

\textbf{SNR}. We defined the signal-to-noise ratio as the ratio between the amplitude of the detection peak and the rms noise of the spectrum. The detected dwarfs have SNR between 3.3 and 45.2. No cut in SNR was applied to the HI sample.\\

We note that, for the dwarfs observed by Arecibo only, we used the W$_{50}$, velocity,  HI flux values, and their respective uncertainties, from the ALFALFA catalog. Moreover, considering the 13 dwarfs detected in both surveys, we find that we measure comparable velocities and W$_{50}$ line widths using the ATLAS$^{3D}$ data as compared to ALFALFA, with an estimated median difference of 1\% and 14\%, respectively. Note that we do not often display the errors on the different properties in the figures, as the error bars would appear smaller than the markers.\\

\begin{figure}
\centering
\includegraphics[width=\linewidth]{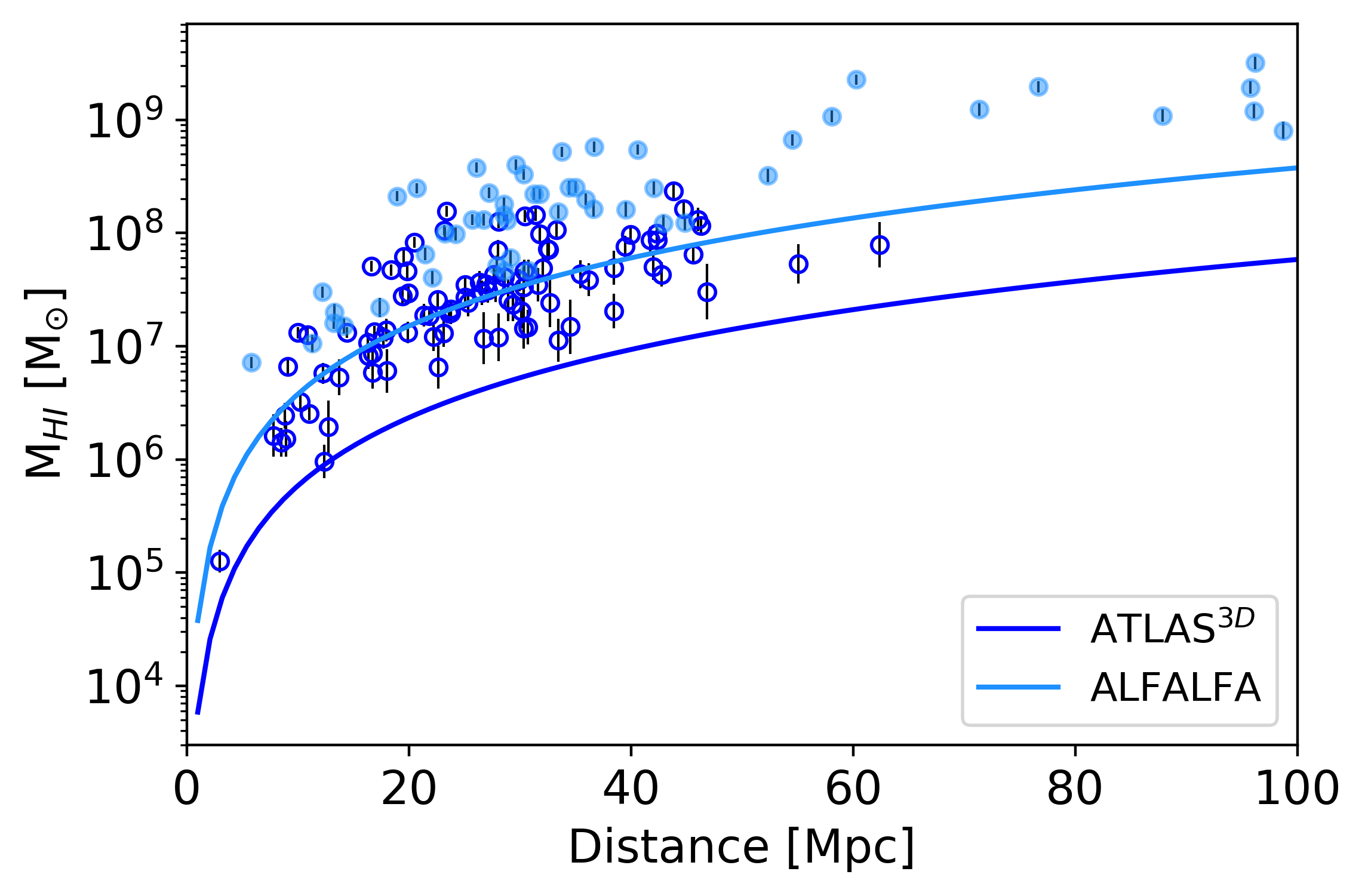}
\caption{HI mass of the detected galaxies as a function of the distance. Solid lines: estimated limit of detection for each survey; blue represents the ATLAS$^{3D}$ HI survey, while light blue represents the ALFALFA survey.}
\label{fig:HImasslim}
\end{figure}

The distribution of selected HI properties is plotted in Figure \ref{fig:HIprop}. It is readily apparent that the ATLAS$^{3D}$ and ALFALFA HI surveys span different distance and HI mass ranges. The difference in the range of distances probed is due to the fact that the ATLAS$^{3D}$ ETGs all have distances $\lesssim$ 45 Mpc, so a maximum velocity corresponding to $\sim$76 Mpc was imposed to the dataset, while ALFALFA can observed galaxies up to $\sim$250 Mpc. The difference in the HI mass distributions is due to the sensitivity of the radio telescopes. In Figure \ref{fig:HImasslim} we show the HI mass calculated for the dwarfs in each survey, as well as the detection limit in HI mass as a function of distance. We estimate the limit of detection using the formula from \citet{Gavazzi2008}:
\begin{equation}
    M_{HI,lim} [M_{\odot}] = SNR \times rms \times W_{50} \times D^2 \times 2.36\times10^5 
\end{equation}
with, for each survey, the minimum SNR and W$_{50}$ of our HI-bearing dwarfs as $SNR$ and $W_{50}$, respectively, the median noise rms of the survey as $rms$, and the HI distance in Mpc as $D$.
We can see that for similar distances, the ALFALFA survey has an HI mass limit about 10 times larger than the ATLAS$^{3D}$ HI survey. Thus, the ATLAS$^{3D}$ HI survey allows us to obtain observations of low HI mass galaxies while the ALFALFA survey extends our detection limit to dwarfs located beyond ATLAS$^{3D}$ distance limit with higher HI masses.

\begin{figure}
\centering
\includegraphics[width=\linewidth]{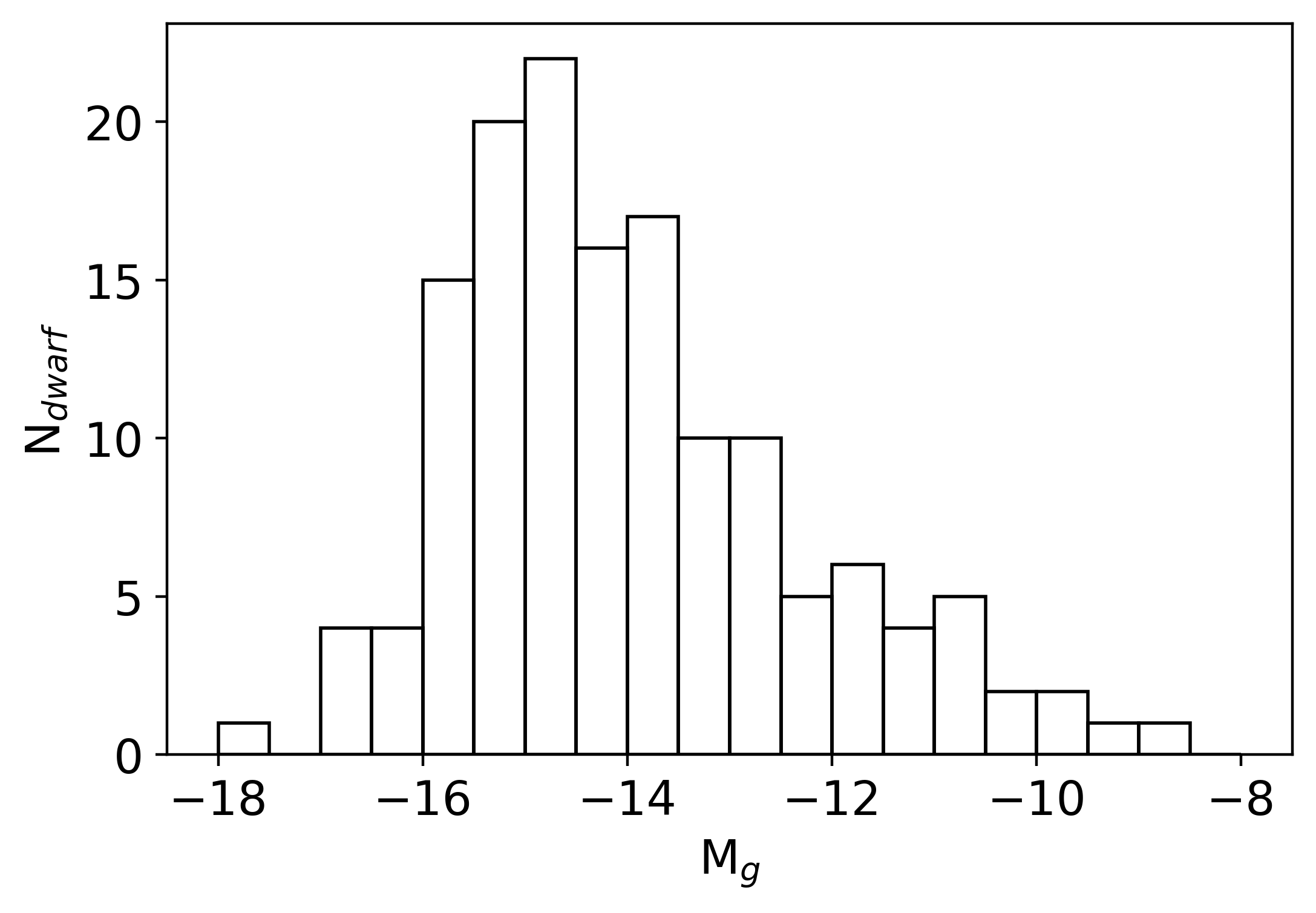}
\includegraphics[width=\linewidth]{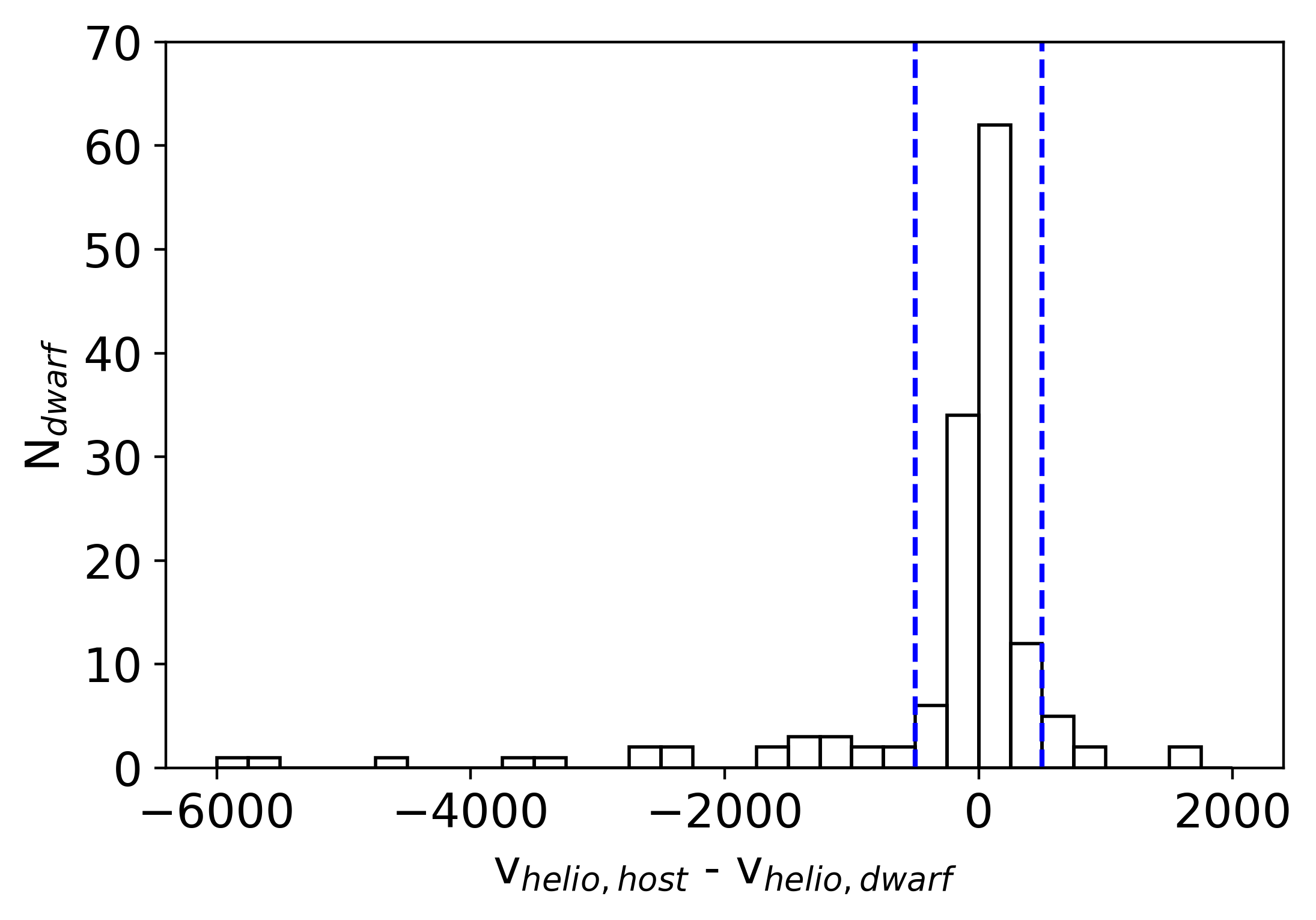}
\caption{Top: the distribution of M$_g$ using HI distances. All the galaxies have M$_g > -18$, confirming that they are dwarfs. Bottom: the difference between v$_{helio}$ of the assumed host ETG and the dwarf. Dashed lines: $\pm$500 km/s cut defining satellites.}
\label{fig:nature_sat}
\end{figure}

Similar to the work from \citet{Habas2020}, where the dwarf and satellite nature of the dwarf sample is probed using available distance measurements (see Section \ref{section:MATLAS_sample}), we investigate the dwarf and satellite nature of the HI detected MATLAS dwarfs based on their measured heliocentric velocities and derived distances. In Figure \ref{fig:nature_sat} we present the distribution of absolute magnitudes M$_g$ computed using the derived distances. The methods used to obtain the dwarfs photometric properties are explained in Section \ref{section:Opt_prop}. All the HI-bearing galaxies have M$_g > -18$. As the B-band magnitudes are systematically fainter than g-band magnitudes in our sample, this confirms their dwarf nature.

\citet{Habas2020} tested the satellite nature of dwarfs considering several assumptions for the hosts. They used the difference of heliocentric velocity between the assumed host and the dwarf, and assumed that the dwarf is a satellite when $\mid\Delta v\mid$~<~500~km/s. This criteria is larger than the typical velocity dispersion observed in Hickson compact groups \citep{Hickson1997} to include groups with possible larger velocity dispersion and the errors on the extracted velocities. We adopt the same $\mid\Delta v\mid$ criteria and verify the assumption that the HI dwarfs are indeed satellites of their assumed host ETG. This confirmation is necessary to study the spatial distribution of the HI dwarfs around their host (see Section \ref{section:environment}). We find that $\sim$ 79\% of the HI dwarfs are satellites of their assumed host, and this fraction increases to 81\% if we consider only dwarfs located in fields with a single ETG. 

As discussed in Section \ref{section:MATLAS_sample}, we may have rejected valid dI galaxies in the original classification process (see \citealt{Habas2020} for more details). We can estimate the number of gas rich dwarfs rejected during the last round of classification by looking at their HI content. For this test we consider all the galaxies not classified as elliptical that were rejected during the last round of classification (280 objects). From this list, 194 are located in the targeted regions of the sky of at least one of the HI surveys. We find an HI detection for 17 and 21 dIs candidates in the ATLAS$^{3D}$ HI survey and ALFALFA survey, respectively. With 3 objects being detected in both surveys, this leads to a sample of 35 HI sources. Based on a visual check of these objects, we rejected 7 HI sources that were actually gas ejected from interacting massive galaxies. To investigate the dwarf nature of the remaining HI sources, 
we derived their HI distances and computed their g-band absolute magnitude M$_g$ using \textsc{source extractor}. Of these, 21 have M$_g$>-18 and thus would be classified as dwarfs, and all have consistent HI masses with a range 6.4 $\lesssim $ log(M$_{HI}$/M$_{\odot}) < 9.8$. We note that the sources not detected in HI can either be background galaxies or galaxies with an HI content too small to be detected by the telescopes. From this result, we can estimate that about 11\% (21/194) of the rejected galaxies are dIs. Thus, we probably missed a few tens of galaxies in our final sample of MATLAS dwarfs.

\begin{figure}
\centering
\includegraphics[width=\linewidth]{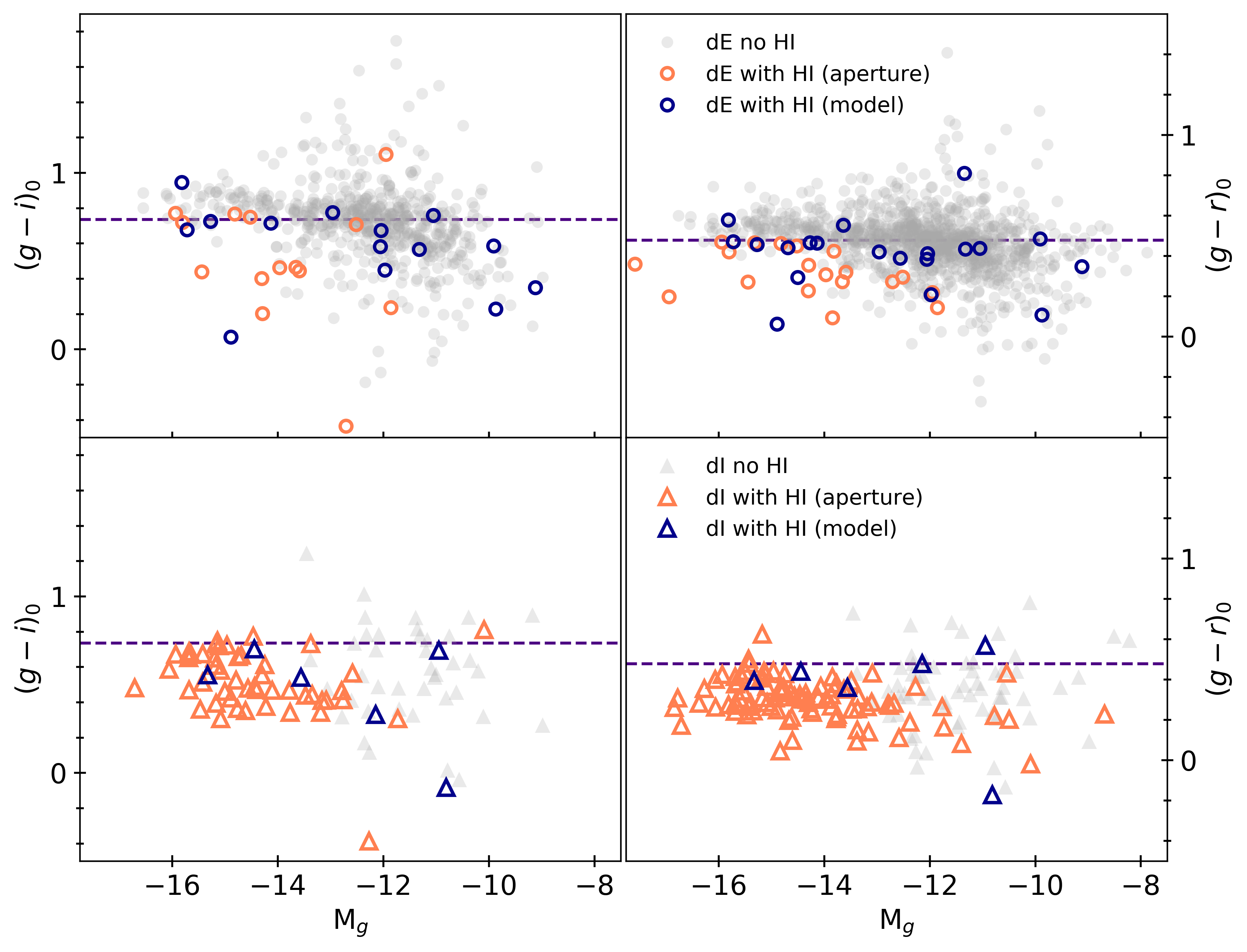}
\caption{The $(g-i)_0$ (left) and $(g-r)_0$ (right) color-magnitude diagrams of the HI MATLAS dwarfs (blue and orange markers) as compared to the ones with no HI detection (grey markers). The colors are computed from \textsc{galfit} models (grey and blue markers) and \textsc{source extractor} aperture photometry (orange markers). Dashed lines: median colors of the sample of MATLAS dwarfs with no HI detection. We separate the dwarfs by dE (top, circles) and dI (bottom, triangles) morphology.}
\label{fig:colors}
\end{figure}

\begin{figure*}
\centering
\includegraphics[width=0.45\linewidth]{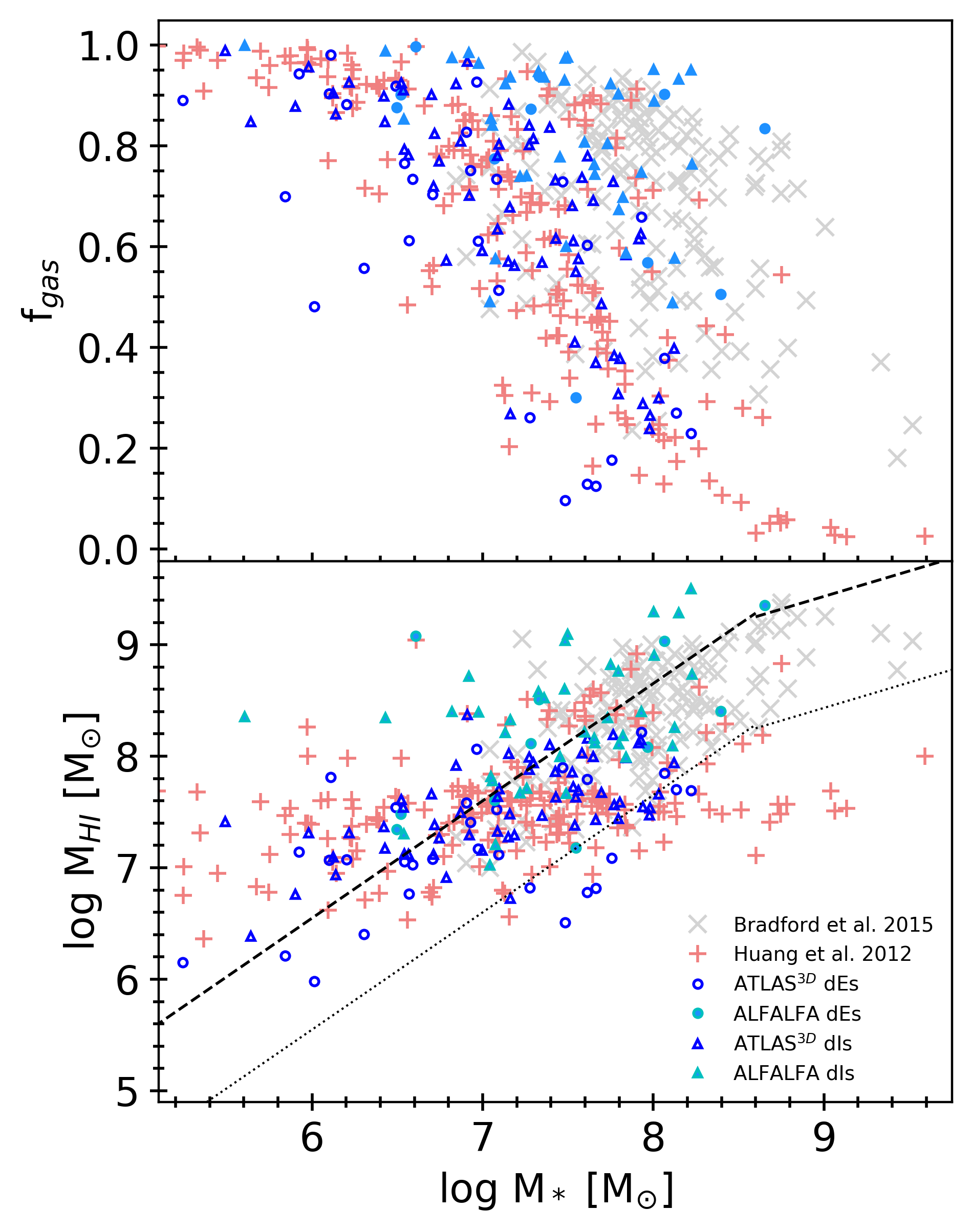}
\includegraphics[width=0.47\linewidth]{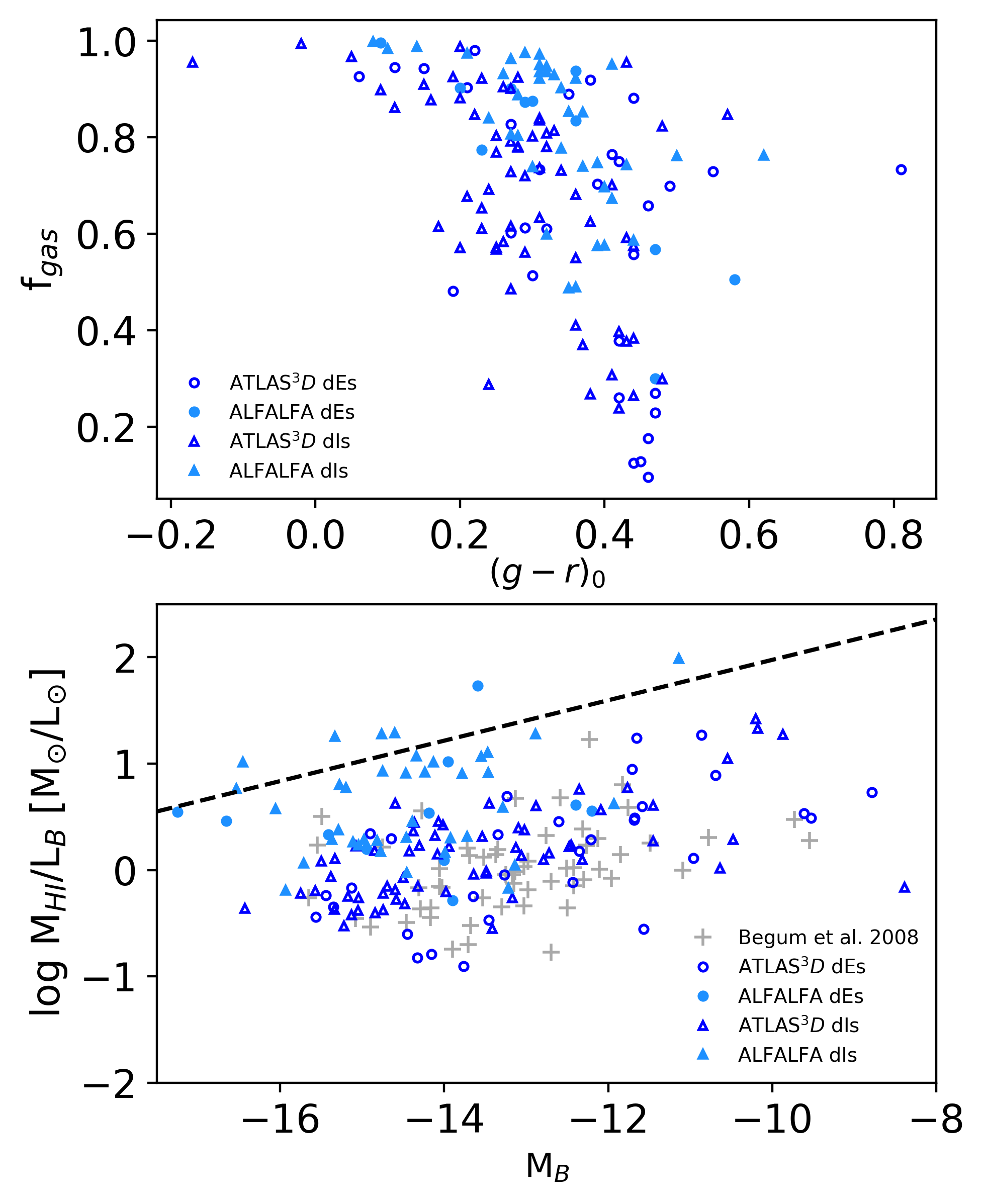}
\caption{Relations between the HI and optical properties of the dwarf galaxies detected in ATLAS$^{3D}$ HI survey (blue markers) and in ALFALFA (light blue markers) as compared to dwarfs (crosses) from \citet{Huang2012} (red), \citet{Bradford2015} (grey), and \citet{Begum2008} (grey). We highlight MATLAS dEs with circles and dIs with triangles. Left: gas fraction (top) and log M$_{HI}$ (bottom) as a function of log M$_{*}$. The relations for galaxies with log(M$_*$/M$_{\odot}$) < 8.6 and > 8.6 from \citet{Bradford2015} are represented with dashed lines, while the dotted lines identify the differences of $-1$ dex from these relations. Right: relations between log M$_{HI}$/L$_B$ as a function of M$_B$ (bottom) and the gas fraction as a function of $(g-r)_0$ (top). The dashed line shows the upper envelope for the HI mass-to-light ratio at a given luminosity from \citet{Warren2007}.}
\label{fig:HIopt}
\end{figure*}

\subsection{Optical properties}
\label{section:Opt_prop}

We measured the photometric properties of the MATLAS HI dwarfs through the use of two methods: 2D surface brightness modeling and \textsc{source extractor} aperture photometry.
Of the 145 HI-bearing dwarfs, 31 have a successful 2D surface brightness model in the g-band \citep{Poulain2021} and we performed aperture photometry on the remaining 114 HI-bearing dwarfs using an aperture size of $\sim$ 3R$_e$ of the dwarf\footnote{R$_e$ estimate, based on the returned \textsc{source extractor} spheroid component effective radius SPHEROID\_REFF\_IMAGE, that we corrected from a deviation observed from the R$_e$ derived with \textsc{galfit} 2D surface brightness modeling for the successfully modeled MATLAS dwarfs.}
, 3R$_e$ being the aperture size returning the closest magnitude value to \textsc{galfit} results.
As the HI mass-to-light ratio is often based on the luminosity in the B-band (e.g., \citealt{Conselice2003,Begum2008,Denes2014}), we converted our g-band to B-band magnitude with the formula $B=g+0.3130\ (g-r)+0.2271$ (Lupton 2005\footnote{\url{http://www.sdss3.org/dr10/algorithms/sdssUBVRITransform.php}.}).
We calculate the stellar mass $M_*$  based on the stellar mass-to-light ratios from \citet{Bell2003}, using the derived $(g-r)$ color, as a larger number of the dwarfs were observed in the r-band than in the i-band. The uncertainty on $M_*$ is estimated using the statistical error on the magnitude and color from either \textsc{galfit} or \textsc{source extractor}, and the error on the distance. The optical properties are available in Table \ref{tab:properties}.

It should be noted that the underlying choice of initial mass function (IMF) assumed in the calculation of the stellar mass will have an impact on the derived stellar masses, especially in the case of low mass galaxies (e.g., \citealt{Huang2012,Boselli2014,Durbala2020}). In particular, \citet{Bell2003} make use of a diet Salpeter IMF while  \citet{Taylor2011} and \citet{Zibetti2009}, which present two other common mass estimates, both opt for a Chabrier (2003) IMF. This variation results in 0.35, 0.37 dex difference for the stellar masses based on \citet{Taylor2011} using $g-i$ color \citep{Habas2020}, and \citet{Zibetti2009} using $g-r$ color, respectively, with the \citet{Bell2003} masses systematically larger. In the remainder of the paper, we take into account this difference when comparing our sample to those using another IMF by correcting the stellar masses of the MATLAS HI-bearing dwarfs according to the IMF favored by the other samples.

\subsubsection{Dwarfs colors}

Most of the optical studies of HI-bearing dwarf galaxies report blue colors caused by recent star formation activity \citep{Grossi2009,Cannon2011,Huang2012,Honey2018}. However, some studies have detected HI in red dEs located in the Virgo cluster \citep{Conselice2003,Hallenbeck2017}.
In Figure \ref{fig:colors}, we show the color-magnitude relation for the Galactic extinction\footnote{We used the reddening values from \citet{Schlafly2011}.} corrected $g-i$ and $g-r$ colors of the MATLAS dwarfs present in the regions of the sky observed by either ALFALFA and ATLAS$^{3D}$ HI survey. If distance measurements are available in the literature, we set those distances to the dwarfs not detected in either HI surveys, otherwise we assume the dwarfs to be located at the distance of the host ETG. The galaxies are grouped by HI content: those with HI measurements are highlighted with colored open markers while those without an HI detection are in grey. Considering both dE and dI morphologies, the HI dwarfs are bluer than the median colors $(g-i)_0 = 0.74$, $(g-r)_0=0.48$ of the dwarfs with no HI detection, with the HI-bearing dIs being bluer (mean $(g-i)_0 = 0.57$ and $(g-r)_0=0.30$) than the HI-bearing dEs (mean $(g-i)_0 = 0.61$ and $(g-r)_0=0.37$).

\subsubsection{Relations between optical and HI properties}
\label{section:scalingrelation}

In Figure \ref{fig:HIopt} we show various relations between the derived HI and optical properties of the MATLAS dwarfs, to test if our HI-bearing dwarfs follow the same relations as other dwarfs of similar masses, and if effects due to the environment can be identified. For this purpose, we selected three samples of dwarfs from the literature based on their size, their provided properties, the range of HI masses studied, and the environments of the dwarfs. The first is the sample of isolated low-mass galaxies from \citet{Bradford2015}, based on the NASA Sloan Atlas catalog \citep{Blanton2011} and observed with the Arecibo and Green Bank telescopes \citep{Geha2006}. The second is the dwarfs sample from \citet{Huang2012} selected from the 40\% ALFALFA catalog \citep{Haynes2011}. The third is the sample of dIs observed by the Giant Metrewave Radio Telescope (GMRT) for the FIGGS (Faint Irregular Galaxies GMRT Survey, \citealt{Begum2008}). The combination of these three samples probe a large range of environments from low to high local densities, as well as HI masses down to $\sim 10^6 M_{\odot}$. The first sample contains only isolated dwarfs, i.e., galaxies with a projected separation from a massive host above 1.5 Mpc. The second sample is composed of a majority of dwarfs located in the field and group environments, and a few tens of galaxies members of the Virgo cluster. In the third sample, the authors selected dwarfs from similar environments to the MATLAS dwarfs, that is, located in the field as well as in groups.

In the left panel, we present the relations between the stellar mass and both the HI mass and the gas fraction, defined as M$_{HI}$/(M$_{HI}$+M$_*$) \citep{Bradford2015}. We compare the MATLAS dwarfs to the samples of \citet{Bradford2015} and \citet{Huang2012}. Note that we computed the stellar masses of the latter sample, as well as for the MATLAS HI sample, using the mass-to-light ratios from \citet{Zibetti2009} and the $u-r$ (available for a larger number of dwarfs than the provided stellar masses based on SED fitting), $g-r$ colors, respectively, to be consistent with the former which provides stellar masses based on a \citet{Chabrier2003} IMF. We see that the gas fraction increases in galaxies with lower stellar masses and that the HI mass increases with the stellar mass in agreement with \citet{Huang2012} and \citet{Catinella2018}. A break in the HI-stellar mass relation between low and high mass galaxies is visible in \citet{Huang2012} and was identified at M$_* = 10^{8.6}$M$_{\odot}$ in the study of \citet{Bradford2015} by fitting power-law functions (broken dashed line in Figure \ref{fig:HIopt}). \citet{Bradford2015} suggest that this break is due to a less efficient star formation activity in dwarfs than in more massive galaxies. This assumption would also apply to the MATLAS HI dwarfs, as they follow the fitted relation for the low-mass galaxies. 

Regarding the environment in terms of the dwarfs, \citet{Bradford2015} found a difference between the HI-stellar mass relation and gas fraction of isolated and non-isolated dwarfs, where only non-isolated galaxies show a deviation from the fitted relation lower than $-1$ dex, and gas fractions below 0.3 for log(M$_*$/M$_{\odot})<9.25$. Focusing on the dwarfs in the samples of MATLAS and of \citet{Huang2012} that fall below the threshold, we remark that they all have f$_{gas}\leq 0.3$. Of the MATLAS sample, 12 show an offset lower than $-1$ dex and all but one are transition-type dwarfs, which appear to be non-isolated galaxies (see Section \ref{section:morphenv}). Moreover, we identify a majority of the deviating dwarfs from \citet{Huang2012} to be members of either the Virgo cluster or nearby groups. From these observations, we can conclude that moderate to high density environments tend to influence the gas content of dwarfs, making them deviate from the HI-stellar mass scaling relation. This could be caused by galaxy harassment or ram-pressure stripping, as suggested by the studies of HI deficient galaxies in groups and clusters (e.g., \citealt{Hess2013,Denes2016}).

In the top right panel of Figure \ref{fig:HIopt} we investigate the fraction of gas as function of the $(g-r)_0$ color of the MATLAS dwarfs. As reported in \citet{Huang2012}, the gas fraction correlates with the color the dwarfs such that the bluer galaxies have the largest gas fraction. In the bottom right panel we show the HI mass-to-light ratio as a function of the absolute magnitude in the B-band. The dashed line represents the upper envelope for the HI mass-to-light ratio, i.e., the minimum fraction of the total baryonic mass to be converted into stars so the galaxy remains gravothermally stable \citep{Warren2007}. As in \citet{Begum2008}, we find that most of the MATLAS HI dwarfs have HI mass-to-light ratios much smaller than the envelop, meaning that they converted a larger quantity of baryons than the one needed for their stability. Moreover, like the \citet{Begum2008}, \citet{Huang2012} and \citet{Bradford2015} dwarfs samples, the majority of the MATLAS HI-bearing dwarfs have a gas fraction larger than 0.5 and thus have a baryonic mass dominated by their HI gas.

The ALFALFA dwarfs with the largest HI-masses seem to behave differently than the rest of the MATLAS dwarfs, showing large fractions of gas and high HI mass-to-light ratios. They are located at the upper envelope of the HI-mass-to-light ratio as some of the more massive galaxies in the study of \citet{Warren2007} and, unlike the other HI dwarfs, converted only the minimum quantity of baryons to remain stable. Moreover, they follow the same HI mass and gas fraction relations with the stellar mass than the most massive galaxies in the sample of \citet{Bradford2015}. Note that most of them are located in the background of the ATLAS$^{3D}$ massive ETGs and LTGs, i.e. at distances above 50 Mpc. We suggest that these dwarfs, unlike the remaining of the MATLAS HI sample, are as efficient in forming stars as their more massive counterparts.

\subsection{Baryonic Tully-Fisher relation}

The power-law correlation observed between the luminosity and the HI line width of galaxies, found by \citet{Tully-Fisher1977}, is known as the Tully-Fisher relation. It is used to determine distances (e.g., \citealt{Tully2000,Springob2007}) as well as to constrain galaxy formation scenario (e.g., \citealt{Desmond2015,Maccio2016}). However, \citet{McGaugh2000} showed that this relation breaks at the low-mass regime, and that a linear relation can be recovered if we use the sum of the gas mass and stellar mass, i.e., the baryonic mass (M$_{bar}$) instead of the luminosity. The resulting relation is called the baryonic Tully-Fisher relation (BTFR).

\begin{figure}
\centering
\includegraphics[width=\linewidth]{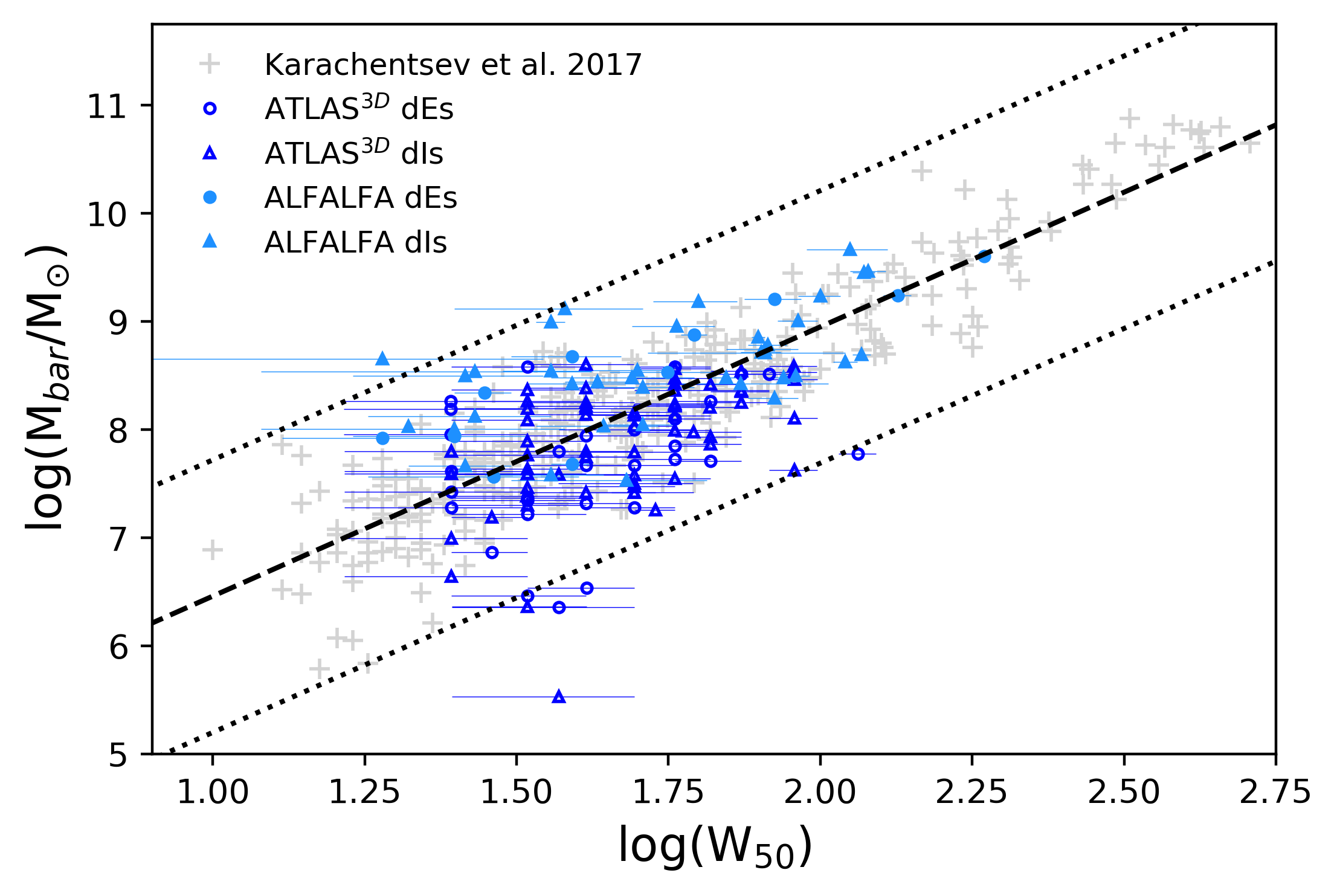}
\caption{BTFR for the MATLAS dwarfs with an HI line detection in the ATLAS$^{3D}$ HI survey (blue markers) and in the ALFALFA survey (light blue markers), as compared to the sample of dwarfs galaxies in the LV (grey crosses) from \citet{Karachentsev2017}. The dashed line represents the linear relation fitted by \citet{Karachentsev2017}. The dotted lines correspond to the maximum scatter of $\pm$1.26 dex for the LV sample. We represent the MATLAS dEs and dIs with circles and triangles, respectively. We use the estimated errors on the velocities for the error bars.}
\label{fig:BTFR}
\end{figure}

We can investigate the consistency of the BTFR obtained for the MATLAS dwarfs as compared to the general trend for dwarfs. As the three catalogs of dwarfs from Section \ref{section:scalingrelation} lack the data to study their BTFR, therefore we choose to compare our sample of HI-bearing dwarfs to the BTFR obtained for the large sample of dwarfs located in the LV from \citet{Karachentsev2017}, taken from the Updated Nearby Galaxy Catalog \citep{Karachentsev2013}. This sample shares similar range of masses to the MATLAS HI-bearing dwarfs with log(M$_{bar}$/M$_{\odot})>5.8$, and has stellar masses derived according to the mass-to-light ratios from \citet{Bell2003}. Similar to \citet{Karachentsev2017}, we define the baryonic mass as $M_{bar} = M_* + \eta\ M_{HI}$ with the factor $\eta = 1.33$ accounting for the Helium abundance in the derivation of the gas mass. The BTFR for the two samples is shown in Figure \ref{fig:BTFR}. The MATLAS HI-bearing dwarfs are consistent with the relation fitted on the LV sample by \citet{Karachentsev2017}. Moreover, we estimated a maximum dispersion from the relation of $\pm$1.26 dex for the LV sample and see that the MATLAS dwarfs show a similar range of scatter, apart from a handful of MATLAS dwarfs that shows a slightly larger dispersion with an absolute value up to 2.35 dex. Among the seven MATLAS dwarfs with a scatter larger than the one observed for the dwarfs from the LV, we see that they all have a significant uncertainty on their W$_{50}$, and that one is the least massive. We note that the scatter of the BTFR is known to increase for dwarfs due to the larger uncertainties encountered in measuring their line width and magnitude (e.g, \citealt{Begum2008b,McGaugh2012}).

\begin{figure}
\centering
\includegraphics[width=\linewidth]{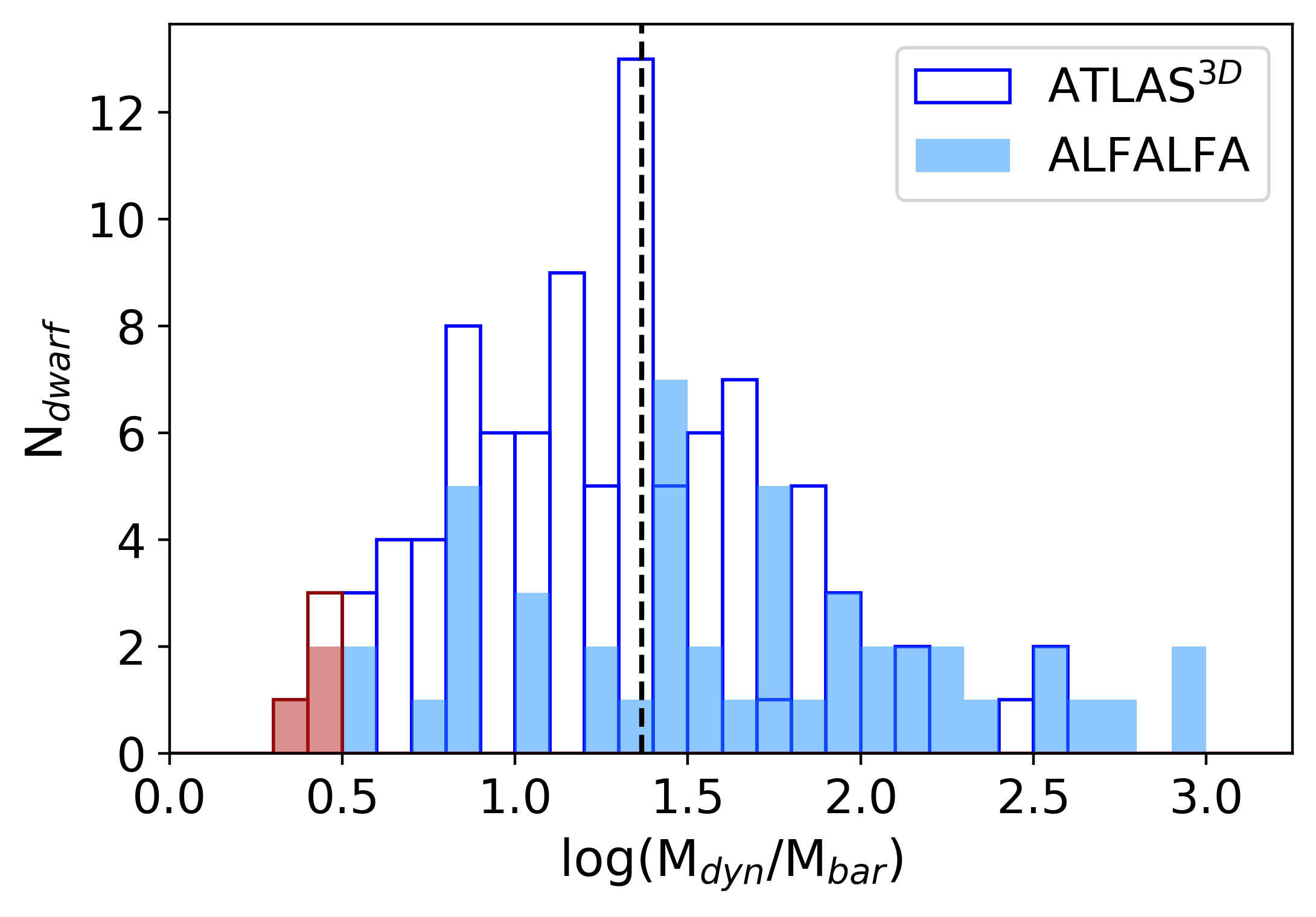}
\caption{Distribution of dynamical-to-baryonic mass ratio of the HI-bearing dwarfs detected in the ATLAS$^{3D}$ HI (empty bars) and the ALFALFA (open bars) surveys. Dashed line: median ratio of the HI dwarf sample. We highlight the dark matter deficient candidates with red colors.}
\label{fig:MdynMbar}
\end{figure}

\subsection{Dynamical masses}

Dwarf galaxies, except those of tidal origin (see Section \ref{section:TDGs}), usually are dark matter dominated (e.g. in the LG; \citealt{Martin2007,Simon2007,Simon2019}). However, in recent years, dwarfs and UDGs with a potential dark matter deficiency have been observed \citep{vanDokkum2018,vanDokkum2019,Emsellem2019,Guo2020}. We can investigate the dark matter content of our HI-bearing dwarfs making use of the dynamical mass.

We estimated the dynamical mass within the HI radius (R$_{HI}$) of the dwarfs following the method from \citet{Guo2020}. They used the formula:
\begin{equation}
    M_{dyn} [M_{\odot}] = 2.31\times 10^{5}\ V_{rot}^2\ R_{HI}
\end{equation}
where $R_{HI}$ is the HI radius in kpc determined using the updated relation observed between the HI diameter (D$_{HI}$) and the HI mass (M$_{HI}$) \citep{Broeils1997} from \citet{Wang2016}:
\begin{equation}
    log D_{HI} = 0.506\ log(M_{HI})-3.293
\end{equation}
and $V_{rot}$ is the rotational velocity in km/s approximated to be $V_{rot} = \frac{W_{20}}{2sin(i)}$. The line width $W_{20}$ is measured at 20\% of the maximum intensity, and $i$ is the inclination angle defined by \citet{Hubble1926} as: 
\begin{equation}
    cos^2(i)=\frac{(\frac{b}{a})^2-q_0^2}{1-q_0^2}
\end{equation}
where $\frac{b}{a}$ is the axis ratio of the dwarf estimated from the Sérsic modeling in the g-band when available (31 galaxies), or from ellipse fitting using \textsc{ellipse} from the \textsc{photutils python} package otherwise, and $q_0$ the intrinsic axis ratio of the dwarfs seen edge-on that we set to 0.2, a typical value used in the literature. 
We note that the value of $q_0$ is likely underestimated for faint dwarfs as it has been shown that they tend to be thicker \citep{Roychowdhury2010,SanchezJanssen2010}.

We find M$_{dyn}$ in the range 7.3 < log(M$_{dyn}$/M$_{\odot}$) $\lesssim$ 11.4 (see Table \ref{tab:properties}), but due to the uncertainty on the galaxies inclination, we may under- or overestimate the dynamical masses. Assuming $q_0=0.6$, the average value found for a sample of faint dIs in \citet{Roychowdhury2010}, the dynamical masses change by a factor of 0.67.

To evaluate the dark matter content of the HI dwarfs, we compute the ratio M$_{dyn}$/M$_{bar}$, for which we show the distribution in Figure \ref{fig:MdynMbar}. With a median dynamical-to-baryonic mass ratio of 23.3, the HI dwarfs seem to be dark matter dominated, as expected. However we observe a wide range with 2.1 $\lesssim$ M$_{dyn}$/M$_{bar}$ $\lesssim$ 879.9 suggesting that some of the galaxies may be dark matter poor. \citet{Guo2020} defined a dwarf as dark-matter deficient when M$_{dyn}$/M$_{bar}<2$. However, a difference of value for $q_0$ from 0.2 to 0.6 involves a difference of $-1$ for the smallest dynamical-to-baryonic mass ratios. For this reason, we consider in our study the dwarfs with M$_{dyn}$/M$_{bar}<3$ as dark matter deficient candidates. This represent a sample of five traditional dwarfs (MATLAS-146, MATLAS-580, MATLAS-1194, MATLAS-1750 and MATLAS-2126) and two UDGs (MATLAS-42 and MATLAS-1824).

\section{Morphology and environment}
\label{section:morphenv}

The most detailed studies of the optical, HI properties, morphology, tidal features and local environment of dwarf satellites have been performed in the LG, for example with the investigation of the gas content of the dwarf population as a function of the distance to the host (either the Milky Way or M31) and the morphology \citep{Grebel2003, Grcevich2009,Mcconachie2012}. However, no such complete analysis has been performed for other satellites systems. Some studies focus on the morphological distribution of dwarfs around more massive galaxies (e.g., \citealt{Ann2017}). Others investigated signs of interactions between dwarfs and their host massive galaxy (e.g., \citealt{Ludwig2012}), or the spatial distribution of the satellites around their host (e.g., \citealt{Mueller2017}). The HI content of dwarfs has also been examined as a function of their location in groups (e.g., \citealt{Hess2013}).
The MATLAS HI-bearing dwarf sample includes a significant fraction of dEs, and about 80\% of the galaxies are estimated to be satellites of nearby massive ETGs. 
In this section, we perform a detail study of the morphology and local environment of the MATLAS HI-bearing dwarf satellites. We focus especially on the dEs, UDGs, TTDs, and interacting galaxies. We also examine the spatial distribution of the dwarfs.

\subsection{Dwarf ellipticals}

Several assumptions have been put forward to explain the observation of HI-bearing dEs.
In the cluster environment, the HI-bearing dEs could be the evolution of an infalling star-forming galaxy into the cluster or could have recently accreted HI gas from an infalling HI cloud \citep{Conselice2003,Hallenbeck2012,Hallenbeck2017}. In low density environments, the HI gas in dEs could result from interactions with a companion or a gas accretion from the intergalactic medium \citep{Grossi2009}.

At most a dozen of HI-bearing dEs have been reported in the outskirts the Virgo cluster \citep{Conselice2003,diSeregoAlighieri2007,Hallenbeck2012,Hallenbeck2017}, and a couple were observed in low density environments \citep{Grossi2009}. With an HI line detection for 42 dEs, our sample is significantly larger than the ones previously reported.
We investigate the variation of the detection rate of the MATLAS HI-bearing dEs with the density of the environment. Considering the MATLAS dEs located in the regions observed by either the ATLAS$^{3D}$ HI survey and the ALFALFA survey, we find a detection rate of 3\% for the low-to-moderate density environments of the MATLAS survey. This is similar to the findings of 2\% in the Virgo cluster \citep{diSeregoAlighieri2007} but lower than the 13\% found in low density environments \citep{Grossi2009}. However, these studies used apparent magnitude limited samples, and selected only galaxies with m$_r$>17.77. Applying a similar cut to the MATLAS dEs we obtain a detection rate of 9.5\%, which is much closer to the one found in low density environments. The detection rate of the MATLAS dE sample decreases toward dwarfs with fainter apparent magnitude. This is probably due to the telescopes detection limits, as fainter dEs are either less massive or too far so their HI content can be detected. Thus, as concluded in \citet{Grossi2009}, we find that the HI-bearing dEs appear to be more numerous in low-to-moderate density environments than in clusters.

Some studies mention a peculiar morphology for HI-bearing dEs \citep{Grossi2009}. As an example, \citet{Hallenbeck2017} argue that some of their studied dEs have a transitional morphology. The likelihood of the MATLAS HI-bearing dEs to be TTDs or the result of galaxy interactions is detailed in sections \ref{section:TTD} and \ref{section:interactions}. 

\subsection{Ultra-diffuse galaxies}

HI-bearing UDGs were recently studied in different environments such as in the field (e.g., \citealt{Leisman2017,Papastergis2017,Janowiecki2019}), in groups \citep{Spekkens2018} or in poor galaxy clusters \citep{Shi2017}. Compared to galaxies of similar HI-mass and environment, these studies report a bluer color for the observed UDGs, a narrower line width and a larger gas fraction.

Of the 59 UDGs in the MATLAS dwarf sample, 51 are located in the regions observed by the ALFALFA and ATLAS$^{3D}$ HI surveys and three have an HI line detection (MATLAS-42,  MATLAS-1337 and MATLAS-1824). These galaxies are located at distances from 33 to 46.5 Mpc with HI masses $8.1\lesssim$ log(M$_{HI})\lesssim 8.3$, line widths $24 <$ W$_{50} \lesssim 36$ km s$^{-1}$, and $(g-r)_0$ color in the range 0.06 -- 0.4. They have M$_{HI}$, W$_{50}$ and $(g-r)_0$ values in the range of those observed for HI-bearing UDGs in other studies (e.g., \citealt{Leisman2017, Kunakaran2020}). We note that, due to their central structures and irregular shapes, not all 145 HI-bearing dwarfs have successful 2D surface brightness modeling and thus an available effective radius and central surface brightness, needed to asses their UDG nature. Moreover, based on the relation between the stellar and HI masses, at least 6 of the 48 UDGs located in the observed regions with no HI detection have stellar masses so low that we neither expect nor detect any HI in either survey. Considering only UDGs with a stellar mass high enough so we might detect HI gas in either the ALFALFA survey or the ATLAS3D HI survey, given their sky location, we estimate that 7\% of them are HI-bearing UDGs. Using the same method for the MATLAS dwarfs which are not UDGs, we find that 10\% are HI-bearing galaxies. Thus, HI-bearing UDGs are slightly rarer in our sample than HI-bearing traditional dwarfs.

We compare the properties of our HI-bearing UDGs to classical dwarfs of similar HI-mass and within the same survey. We find that the UDG line widths are smaller than the median line width of the HI dwarfs. Only one UDG in each survey has an available observation in the r and i bands and thus an available stellar mass, $(g-r)_0$ and $(g-i)_0$ colors. The UDG detected in the ATLAS$^{3D}$ HI survey has a gas fraction larger and bluer colors than all the ATLAS$^{3D}$ dwarfs of similar HI-mass, while the UDG detected in the ALFALFA survey has a smaller gas fraction and redder color than the median values of the dwarfs.
We note that these two UDGs are dark matter deficient candidates with a dynamical-to-baryonic mass ratio of 2.8 and 2.9 for MATLAS-1824 and MATLAS-42, respectively.

\subsection{Transition-type dwarfs}
\label{section:TTD}

TTDs have properties of both dEs and dIs, generally showing central irregular star-forming regions and an outer region composed of an older stellar population associated with smooth elliptical external isophotes \citep{Dellenbusch2008,Koleva2013}. In clusters, these dwarfs are thought to correspond to a transition phase in the transformation of LTGs into quescient elliptical dwarfs when going through ram-pressure stripping during their infall \citep{Boselli2008}. In lower density environments, they could be the result of an internal interstellar medium instability or a galaxy-galaxy interaction, that would explain the central blue star-forming regions often observed \citep{Dellenbusch2008}. They also could be dIs or dEs going through episodic star-forming or gas infall, respectively \citep{Koleva2013}.

We investigated the morphology of the dwarfs as a function of the HI mass-to-light ratio $M_{HI}/L_B$. Although TTDs were not identified during the initial classification of the MATLAS dwarf sample, they can be identified as active star-forming galaxies with a mass-to-light ratio $0.1 \lesssim M_{HI}/L_B \lesssim 0.5$, while dEs usually are gas poor with ratios lower than 0.1 (in the LG, e.g.,  \citealt{Mcconachie2012}) coupled with an absence of star formation activity and the dIs are actively forming stars with a ratio greater than 1 \citep{DaCosta2007}. In our sample, 17 dwarfs (8 dEs and 9 dIs) have an HI mass-to-light ratio consistent with a transitional morphological type. These dwarfs all show an elliptical shape with inner star-forming regions (see Figure \ref{fig:colorcutouts}).

A number of galaxies in the sample have visual morphologies that do not correspond with the morphology expected from their HI mass-to-light-ratios. For example, 20 dwarfs identified as dIs have $0.5 < M_{HI}/L_B < 1$. These dwarfs all show an elliptical shape as well as internal structures (e.g. MATLAS-1750, MATLAS-519), large star-forming regions or signs of dwarf-dwarf interactions (e.g. the shells of MATLAS-602 discussed in Section \ref{section:interactions}). On the other hand, 28 dwarfs identified as dEs have $M_{HI}/L_B > 1$. Finding dEs with a high HI mass-to-light ratio is not unexpected, as some have been observed in the outskirts of clusters \citep{Conselice2003,Buyle2005}. Three of these are tidal dwarf candidates (MATLAS-1824, MATLAS-1322 and MATLAS-1830; see Section \ref{section:interactions}), explaining their high gas content that might originate from massive galaxies. Three other dEs (MATLAS-696, MATLAS-720 and MATLAS-121) have a transitional morphology with an outer elliptical envelope with blue knots of star formation in the interior. They might be transitional dwarfs with an HI gas content as rich as the dIs, as already observed in \citet{Koleva2013}. We note that some of the remaining gas-rich dEs appear to be blue (e.g. MATLAS-714 or MATLAS-1411) and resemble some of the gas-poor dIs previously mentioned. These dEs, together with the gas-poor dIs are difficult to classify, especially without information about the star formation rate, as they have properties of both ellipticals (outer isophotes shape) and irregulars (e.g. a blue color, large star-forming regions).

\begin{figure*}
\centering
\begin{subfigure}{0.3\linewidth}
 \centering
 \includegraphics[width=\linewidth]{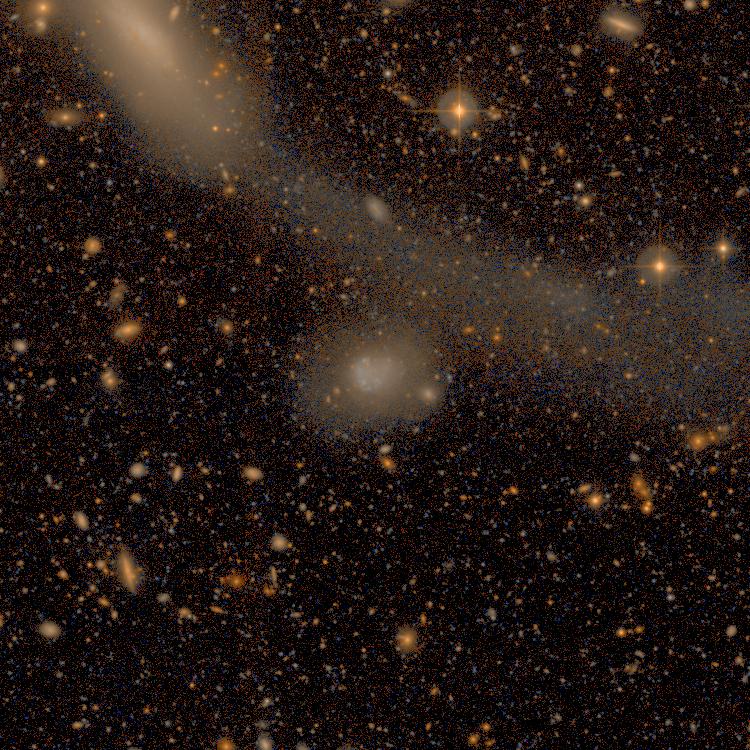}
 \caption{}
\end{subfigure}
\begin{subfigure}{0.3\linewidth}
 \centering
 \includegraphics[width=\linewidth]{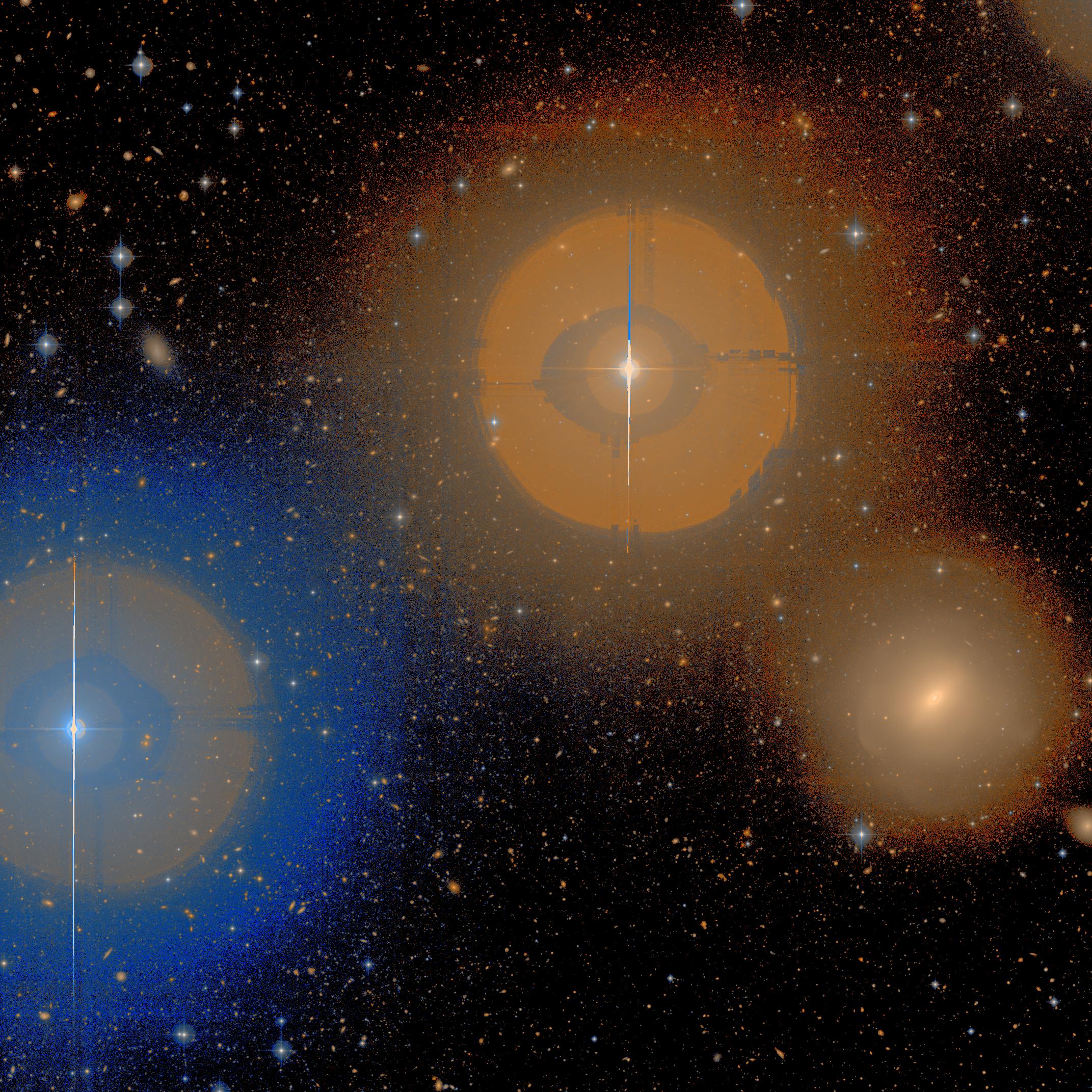}
 \caption{}
\end{subfigure}
\begin{subfigure}{0.3\linewidth}
 \centering
 \includegraphics[width=\linewidth]{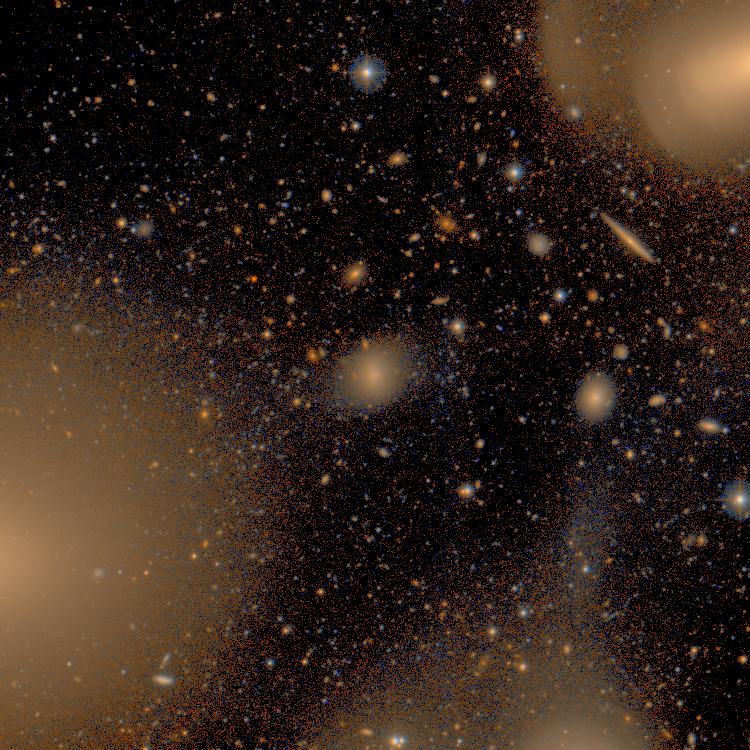}
 \caption{}
\end{subfigure}
\caption{The three newly identified TDG candidates: a) MATLAS-1180 (center), connected to a tidal tail extending from an interacting galaxy visible in the top-left corner; b) MATLAS-947 (top-left corner), tidally connected to the ETG NGC3610 (bottom-right corner) showing shell features; c) MATLAS-1322 (center), with extended isophotes towards interacting galaxies on the bottom-left and top-right corners. The images are 7\arcmin\ $\times$ 7\arcmin\ for a) and c), and 20\arcmin\ $\times$ 20\arcmin\ for b), with North up and East left. RGB images were produced with the help of the \textsc{Astropy} package, based on the method from \citet{Lupton2004}.}
\label{fig:TDG}
\end{figure*}

\begin{figure}
\centering
\begin{subfigure}{0.4\linewidth}
 \centering
 \includegraphics[width=\linewidth]{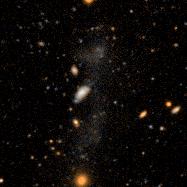}
 \caption{}
\end{subfigure}
\begin{subfigure}{0.4\linewidth}
 \centering
 \includegraphics[width=\linewidth]{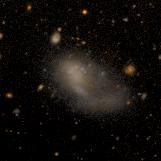}
 \caption{}
\end{subfigure}
\begin{subfigure}{0.4\linewidth}
 \centering
 \includegraphics[width=\linewidth]{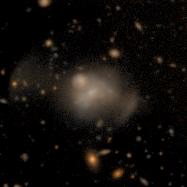}
 \caption{}
\end{subfigure}
\begin{subfigure}{0.4\linewidth}
 \centering
 \includegraphics[width=\linewidth]{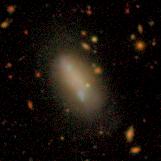}
 \caption{}
\end{subfigure}
\caption{Examples of dwarf interaction features: a) Low surface brightness pair with MATLAS-1783 (North of center) and MATLAS-1784 (South of center) and a central bridge extending in the center of the image from MATLAS-1783 to MATLAS-1784; b) The antennae-like system MATLAS-546; c) MATLAS-602, showing faint shell features on both sides; d) MATLAS-824, also showing faint shell features on both sides. The images are 1.75\arcmin\ $\times$ 1.75\arcmin\ for a) and c), and 1.5\arcmin\ $\times$ 1.5\arcmin\ for b) and d), with North up and East left. RGB images were produced with the help of the \textsc{Astropy} package, based on the method from \citet{Lupton2004}.}
\label{fig:tidalfeatures}
\end{figure}

\subsection{Galaxy interactions}
\label{section:interactions}

Fine structures such as tidal tails or shells are often by-products of galaxy mergers. In the context of dwarf galaxies, two types of galaxy interactions are observed: the interaction between a dwarf and a massive galaxy, and dwarf-dwarf interactions. In the case of a dwarf-massive galaxy interaction, the dwarf can either be the result of a merger event or a tidal interaction between two massive galaxies, the so-called tidal dwarf galaxies (TDGs), or be a future minor merger of its host massive galaxy. With its deep imaging, the MATLAS survey allows us to identify low surface brightness features of galaxies. In this section, we discuss the observed fine structures around the HI-bearing dwarfs for the two types of galaxy interactions.

\subsubsection{Tidal dwarf candidates}
\label{section:TDGs}

TDGs are formed from the gas and stellar material ejected in the interstellar medium during the interactions between two massive galaxies \citep{Duc2000}. TDGs, especially as they age, show similar structural properties to classical dwarfs. However, their properties differ in three main ways: they lack dark matter, have high metallicities \citep{Hunter2000,Duc2007,Sweet2014,Lelli2015} and have no globular clusters \citep{Jones2021}. \citet{Duc2014} previously identified seven, likely old, TDGs around ATLAS$^{3D}$ ETGs showing fine structures. Among these dwarfs, four have an HI line detection in our study (MATLAS-785, MATLAS-1750, MATLAS-1824 and MATLAS-1830).  Of the other three candidates, one is not in the MATLAS dwarf sample due to its very low surface brightness, while the other two were excluded as potential background galaxies. We also identified three new TDG candidates (see Figure \ref{fig:TDG}): the first (MATLAS-1180) is located near the end of the tidal tail of a galaxy that seems to interact with the ETG NGC4111, the second (MATLAS-947) has extended isophotes oriented towards the ETG NGC3610, and the third (MATLAS-1322) has outer isophotes twisted towards the massive interacting galaxies NGC4281 and NGC4270. Therefore, $\sim$5\% of our HI-bearing galaxies are possible TDGs. 
Of the TDGs candidates in \citet{Duc2014}, two have effective radius and central surface brightness measurements consistent with a UDG classification. These two galaxies are in our sample of HI-bearing dwarfs. However, we note that one of these UDGs (MATLAS-1830) is not in the subsample of 59 UDGs considered in this work due to the absence of successful 2D surface brightness modeling caused by its very low surface brightness.
Of the TDG candidates, three have low dynamical-to-baryonic mass ratios as compared to the whole HI dwarf sample, including two dark matter deficient candidates, which is consistent with these galaxies being dark matter poor.

\subsubsection{Dwarf merger candidates}
\label{section:tidalfeatures}

Understanding the role of the merging process in dwarf galaxy evolution is important as they might be the progenitors of some dEs \citep{Kazantzidis2011,Graham2012,Toloba2014,Tarumi2021}.
The improvement of imaging facilities this last decade has made the observation of low surface brightness structures easier \citep{Abraham2014,Duc2014,Bilek2020}. This includes the faint structures resulting from dwarf galaxies interactions. As a consequence, there have been an increasing number of studies focusing on dwarf galaxy interactions and mergers  \citep{Rich2012,Paudel2015,Annibali2016,Pearson2016,Paudel2017,Paudel2018}. These works identify different structures such as tidal tails, plumes, bridges, stellar streams, shells or merged pairs of galaxies.

Similar to the classification system of \citet{Paudel2018}, we define four categories of interaction features: pairs, Antennae-like systems, shells and extended isophotes.
These categories likely represent different merging stages, e.g., with pairs or Antennae-like systems occurring in early stages, and shells being visible in late-stages.
In total, $\sim$ 10\% of the HI-bearing galaxies show signs of interactions. We display examples of the observed features in Figure \ref{fig:tidalfeatures}, while all the dwarf merger candidates are visible in Figure \ref{fig:mergers}. 

We report six possible pairs of interacting galaxies involving the HI-bearing dwarfs: the two HI-brearing dwarfs MATLAS-595 and MATLAS-596 interacting together, MATLAS-340 which shows extended ispohotes and an HI content overlapping both the dwarf and a nearby galaxy, MATLAS-1411 which is possibly interacting with a nearby galaxy with a tidal tail, MATLAS-1783 with a visible bridge connected to MATLAS-1784 and MATLAS-2160 that shows twisted extended isophotes that might be connected to the extended isophotes of MATLAS-2158. We find also a dwarf galaxy (MATLAS-546) whose shape resembles the Antennae system. 

We observe shell-like features in three dwarfs: MATLAS-824, MATLAS-602, MATLAS-621. The first one exhibits symmetric structures, suggesting a merger event between two dwarfs of similar mass \citep{Paudel2017}. The second has asymmetric shell features, implying a different mass for the two dwarf mergers \citep{Paudel2018}. The last shows internal shell-like structures as well as a tidal tail, possible signs of a ongoing merger event. 

We note three dwarfs showing extended isophotes with no clear origin. MATLAS-124 has extended isophotes and could be two dwarf galaxies interacting with a central bridge. MATLAS-580 has two central bright point sources, but its extended isophotes can also be contaminating Galactic cirrus. Finally MATLAS-447 has extended isophotes but is located in a stellar halo, making difficult to observe other nearby low surface brightness structures. 

\begin{figure}
\centering
\includegraphics[width=\linewidth]{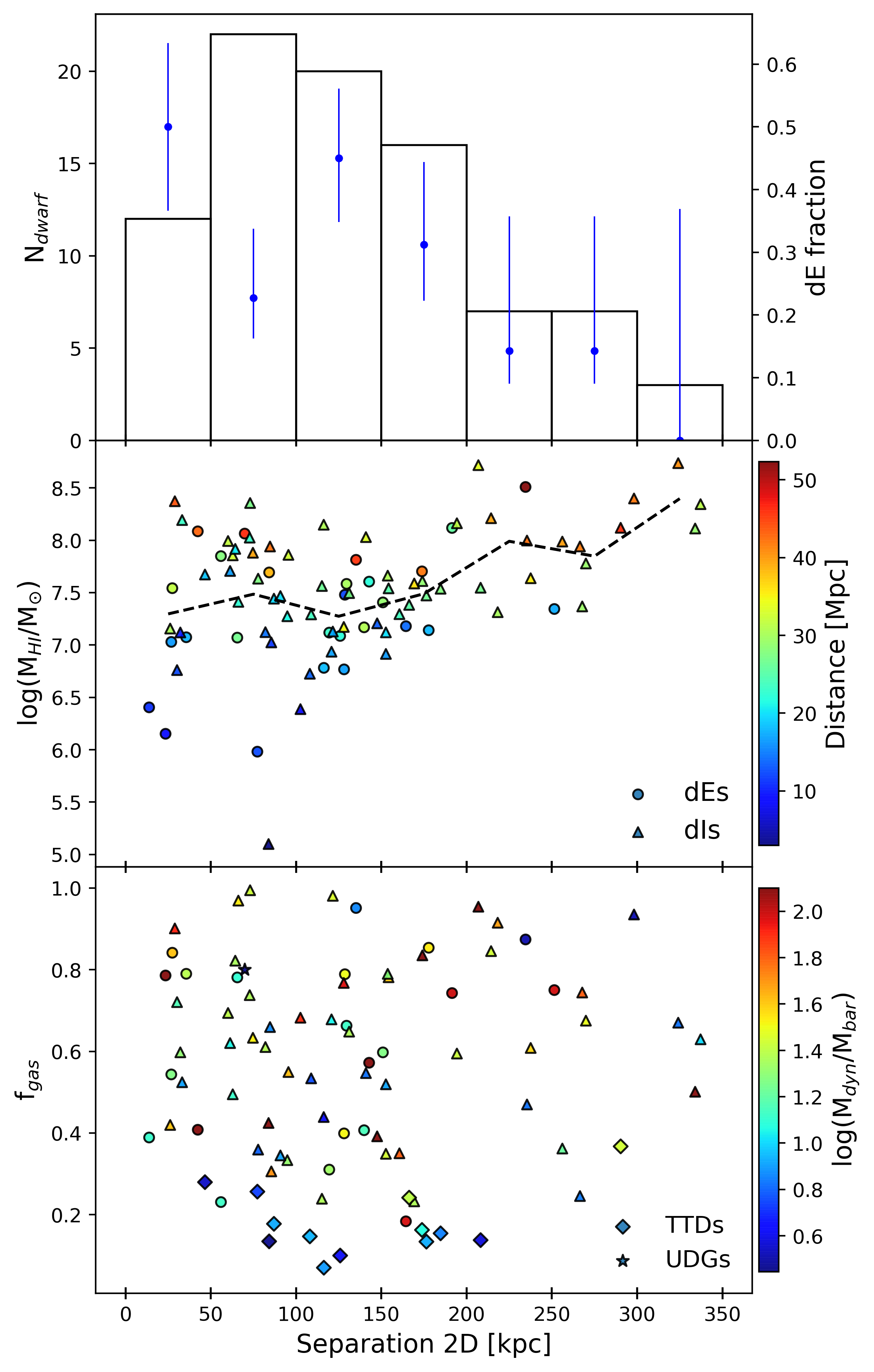}
\caption{HI-bearing dwarf satellites projected separation from the host for the subsample of 50 ETGs (top) as a function of the HI mass (middle) and gas fraction (bottom). We show the fraction of dEs with blue dots and error bars (top), defined as the 1$\sigma$ binomial confidence intervals. The colorbars indicate the distance of the dwarf in Mpc (middle) and the dynamical-to-baryonic mass ratio (bottom). The dashed line represents the running average of the HI mass per bin of width 50 kpc (middle). We highlight the dEs with circles, the dIs with triangles, and added the TTDs as diamonds, and UDGs as stars to the bottom plot.}
\label{fig:spatialdist}
\end{figure}

\subsection{Spatial distribution}
\label{section:environment}

The HI content of galaxies appears to be correlated to their environment and morphology.
The HI mass decreases with an increase of the local density for galaxies of similar stellar mass \citep{Denes2014}. 
In the Local Group, dwarfs located within $\sim$270 kpc of their host (Milky Way or Andromeda) have M$_{HI}$ < 10$^5$ M$_{\odot}$ and a dE morphology, while TTDs and dIs are found at larger distances with higher HI masses \citep{Grebel2003,Grcevich2009}.

\subsubsection{Separation to the host}

We investigate the projected separation around the host ETG, of the dwarfs with HI content. We consider here only fields with a single ETG for which at least one HI satellite is confirmed (see Section \ref{section:HIprop}), which restricts the sample to 50 ETGs with distances from 11 to 45 Mpc, as taken from the ATLAS$^{3D}$ ETG catalog \citep{Capellari2011}. The variation of the HI mass and gas fraction of the satellites are shown as a function of the projected distance from the host in Figure \ref{fig:spatialdist}. We observe, on average, an increase of the HI mass with the projected distance to the host and also a distance limit of $\sim$100 kpc for the detection of dwarfs with log M$_{HI}$/M$_{\odot}$ < 6.5. This result is consistent with the findings of \citet{Denes2014} and \citet{Bouchard2009}, where at fixed stellar mass, the HI mass of dwarfs increases towards lower local density environments, as well as with observations in the LG \citep{Grcevich2009}, where the HI masses of the dwarfs increases with the distance to the host. However, unlike the LG, this trend is not obvious for the gas fraction. This can be due to a detection limit, as this relation is not clearly visible without observations of dwarfs with f$_{gas}<0.1$ in the LG \citep{Grcevich2009,Mcconachie2012}.

Cosmological simulations from \citet{Jackson2021} suggest that dark matter deficient dwarfs are produced via tidal interactions between dwarfs and massive galaxies, and thus the majority of dark matter deficient dwarfs should be located within $\sim$150 kpc of the host ETG. Looking at the fraction of the dwarfs located below and above 150 kpc with a very low dynamical-to-baryonic mass ratios as compared to the whole MATLAS sample, defined here as log(M$_{dyn}$/M$_{bar}$) < 0.8, we observe that $\sim73\%$ of the satellites with a low ratio are located below 150 kpc from the host. However, we note that we do not observe a trend for the dark matter poor satellites to be located closer to the host than dark matter dominated dwarfs.

We focus on the morphology of the dwarfs as function of the projected separations. We can see in Figure \ref{fig:spatialdist} that the fraction of dEs seems to decrease with larger separation, while the TTDs are located at intermediate distances. These observations are in agreement with the spatial distribution of the LG dwarfs \citep{Grebel2003,Grcevich2009}. Note that, unlike the LG, we observe a large fraction of dIs located close to the host, with about 50\% of the HI-bearing dwarfs being dIs at a projected distance below 50 kpc. \citet{Ann2017} investigated the fraction of dEs and dIs as a function of the distance to the host and its morphology (either an ETG or a LTG) and found that $\sim$40\% of the satellites of ETGs located within 0.2 virial radius are dIs. This finding can explain the obtained fraction of dIs for the MATLAS sample. Only one HI-bearing UDG (MATLAS-1824) is satellite of an ETG from this subsample, the other two being likely located in groups. This UDG is relatively close to its host and show a very low dynamical-to-mass ratio, in agreement with its tidal origin (see Section \ref{section:TDGs}.)

\subsubsection{Local density}

\begin{figure}
\centering
\includegraphics[width=\linewidth]{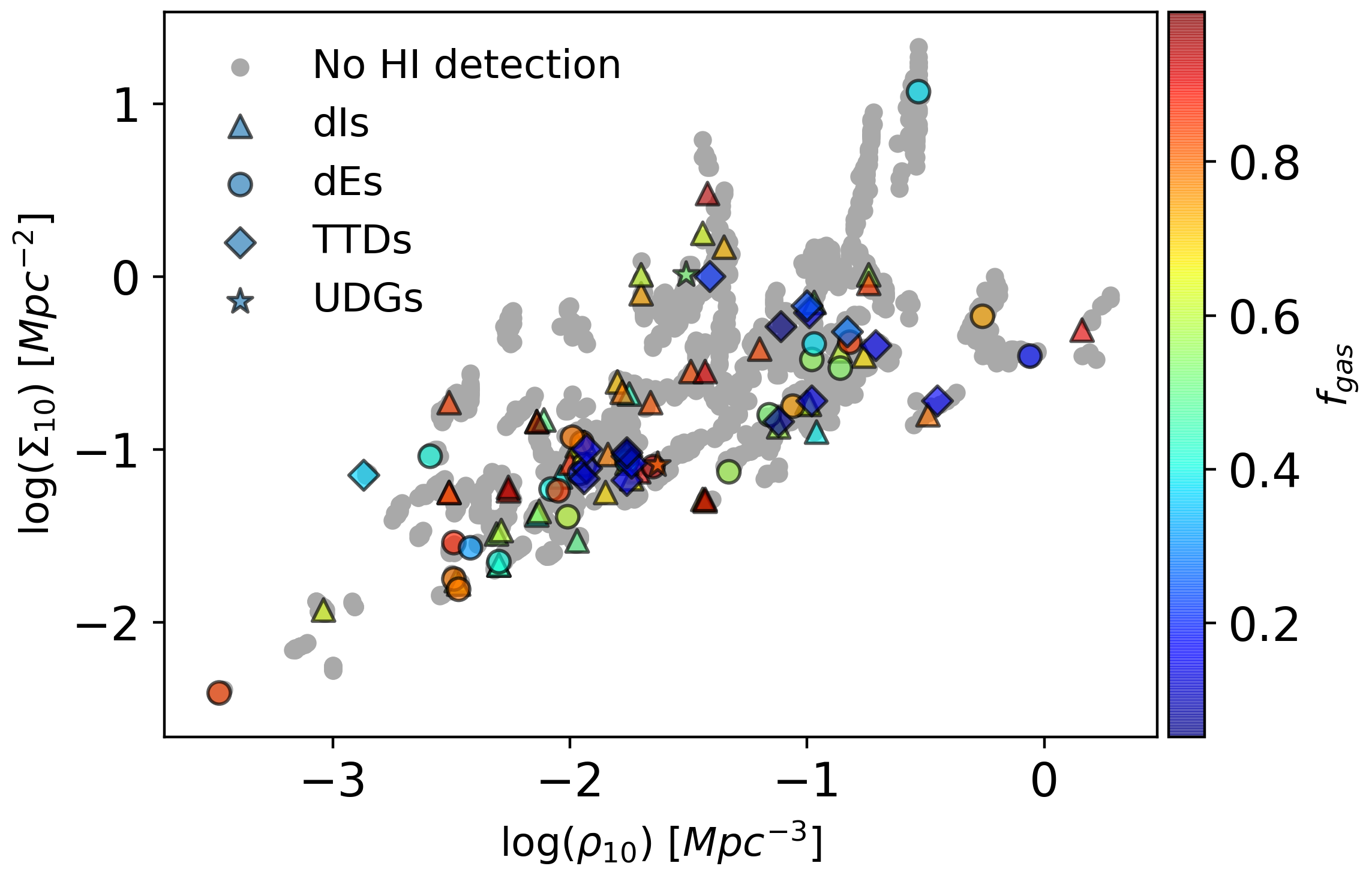}
\caption{The local volume density $\rho_{10}$ as a function of the local surface density $\Sigma_{10}$. We illustrate the local densities probed by the MATLAS survey with the dwarfs with no HI detection (grey dots). We show the HI-bearing dwarfs divided per morphology: the most gas-rich dIs with M$_{HI}$/L$_B$>1 (triangles), the dEs (circles), the TTDs (diamonds) and the UDGs (stars). We indicate the gas fraction of the HI-bearing dwarfs with the colorbar.} 
\label{fig:localdensity}
\end{figure}

We examine possible effects of the local density on the gas content and the morphology of the HI-bearing dwarfs. We make use of two different parameters to estimate the local density of the dwarfs: the local volume density and the local surface density defined as $\rho_N = 3N/4\pi r^3$ and $\Sigma_N = N/\pi r^2$, respectively, with N the N-nearest galaxies and r the radius enclosing these galaxies in Mpc \citep{Cappellari2011b}. We consider for our study the 10 nearest ETG and LTG neighbors of each dwarf from the ATLAS$^{3D}$ catalog \citep{Cappellari2011}, assuming that the dwarf is located at the distance of its assumed host. To ensure the consistency of the local density estimates, we select the HI-bearing dwarfs confirmed to be satellites of their assumed host ETG. We define four subsamples based on the morphology and HI mass-to-light ratios of the dwarfs: UDGs, TTDs (see Section \ref{section:TTD} for definition), dEs, and gas-rich dIs with M$_{HI}$/L$_B$>1. We show in Figure \ref{fig:localdensity} log($\rho_{10}$) as a function of log($\Sigma_{10}$) for the different morphologies. The TTDs and UDGs appear to populate intermediate densities while the gas-rich dEs and dIs are visible along the full range of densities probed by the MATLAS dwarfs.

We computed the running average of the gas fraction using three bins of log($\rho_{10}$) in the range $-3.0$ to 0.0. We find that the gas fraction slightly decreases with increasing local volume density with a mean of 0.63$\pm$0.20, 0.59$\pm$0.28 and 0.48$\pm$0.24 for each bin. Note that we observe a similar trend when using the local surface density as we obtain average gas fractions of 0.62$\pm$0.18, 0.59$\pm$0.26 and 0.54$\pm$0.27 considering three bins in the range $-2.5$ -- 0.5. This trend is consistent with the environment dependence found by \citet{Denes2014}.

\section{Conclusions}

We have studied the HI content of the optically identified MATLAS dwarfs, located in low to moderate density environments. We used radio observations from the ATLAS$^{3D}$ HI survey, performed by the WSRT, coupled with the final realease of the extragalactic HI sources catalog of the ALFALFA survey, based on radio data from Arecibo. These surveys are complementary, as the WSRT detects objects with smaller HI masses than Arecibo and the ALFALFA survey observes galaxies at farther distances than the ATLAS$^{3D}$ HI survey. Of the 1773 sources located in the regions targeted by either surveys, 8\% (145) have an HI line detection. This sample of HI-bearing dwarfs includes 81 dwarfs with previously unreported HI detection. The majority of the dwarfs show irregular morphology (103 or 71\%), while 29\% (42) are ellipticals, the largest sample of HI-bearing dwarf ellipticals (dEs) to date. Based on their structural properties, 2\% (3) are classified as UDGs. We derived their HI properties and compared them with their optical properties based on the Sérsic modeling \citep{Poulain2021} and aperture photometry. We found that 79\% are satellites of their assumed host ETG. 

Based on the derived properties, we find that the HI-bearing dwarfs have bluer colors than the median color of the dwarfs with no HI detection. Moreover, the scaling relations between the HI and optical properties (M$_{HI}$, M$_*$, f$_{gas}$, color and M$_B$) of the MATLAS HI-bearing dwarfs are consistent with the ones for dwarfs of similar mass and local environment. We observe that the MATLAS dwarfs with f$_{gas} < 0.3$ tend to deviate from the HI-to-stellar mass relation of isolated dwarfs and are located in moderate local densities, suggesting an influence of the environment on the HI gas content. Most of the HI-bearing dwarfs show a gas fraction and an HI mass-to-light ratio typical of a galaxy with a baryonic mass dominated by the HI gas. Note that the dwarfs with high HI mass detected by ALFALFA, show a very high gas fraction and HI mass-to-light ratio as compared to the other dwarfs. We suggest this difference to be linked to the galaxy star-forming activity, with these dwarfs being as efficient in forming stars as more massive galaxies, unlike the other HI-bearing dwarfs. In addition to the described scaling relations, the MATLAS HI-bearing dwarfs show a BTFR consistent with the general trend for dwarfs in the LV. Based on the distribution of dynamical-to-baryonic mass ratios, the HI-bearing dwarfs are dark matter dominated, but we identified 5\% (7) dark matter deficient candidates. 

We investigated the morphology and environment of the HI-bearing dwarfs by looking at their HI mass-to-light ratios, the presence of galaxy interaction features and the spatial distribution of HI-bearing dwarfs satellites around their host ETGs. We report several findings. Focusing on the dEs, we estimate a detection rate of HI-bearing dEs in the MATLAS sample of 9.5\%, suggesting as in \citet{Grossi2009} that HI-bearing dEs are more present in low-to-moderate density environments than in cluster. We compared UDGs to traditional dwarfs and find that, considering galaxies with location and stellar masses such that they can be detected in at least one of the surveys, there is a slightly smaller proportion of HI-bearing UDGs (7\%) than HI-bearing traditional dwarfs (10\%), and the UDGs have a smaller line width than the median value of the traditional dwarfs but have no tendency to be bluer or have a larger gas fraction. By investigating the morphology of the MATLAS HI-bearing dwarfs, we find that 12\% (17) have HI mass-to-light ratios and morphological properties consistent with a transitional-type morphology, 2\% (3) dEs are TTD candidates as gas-rich as dIs, 5\% (7) show in their environment interaction features typical of massive galaxy interactions suggesting that they are possibly tidal dwarf galaxies, and 10\% (14) possess low surface brightness structures characteristic of galaxy interactions such as shells, tidal tails or bridges. By studying the local environment, we observe an increase in the running average of the observed HI mass as a function of the projected separation to the host ETG, while the fraction of HI-bearing dEs decreases with the separation to the host ETG, and the TTDs are located at intermediate distances. Lastly, the gas fraction of the HI-bearing dwarfs depends on the local density, with the average gas fraction of the satellites increasing toward low density environments.

In summary, this study provides a large optically and HI selected sample of dwarf galaxies, including UDGs, outside galaxy clusters. The new generation telescopes in HI (e.g., Apertif, ASKAP, MeerKAT) and optical (e.g., Euclid, LSST) will provide a gain in both spatial coverage and HI sensitivity allowing us to better quantify the role of the environment on the gas content and morphology of dwarfs located in the LV and beyond.

\begin{acknowledgements}
The authors thank the referee for the constructive comments that helped to improve this manuscript.
Based on data obtained with the Westerbork Synthesis Radio Telescope (WSRT). The WSRT is operated by ASTRON (Netherlands Institute for Radio Astronomy) with support from the Netherlands Foundation for Scientific Research (NWO). Based on observations obtained with MegaPrime/MegaCam, a joint project of CFHT and CEA/IRFU, at the Canada-France-Hawaii Telescope (CFHT) which is operated by the National Research Council (NRC) of Canada, the Institut National des Science de l'Univers of the Centre National de la Recherche Scientifique (CNRS) of France, and the University of Hawaii. This work is based in part on data products produced at Terapix available at the Canadian Astronomy Data Centre as part of the Canada-France-Hawaii Telescope Legacy Survey, a collaborative project of NRC and CNRS. The authors acknowledge the work of the entire ALFALFA collaboration team in observing, flagging, and extracting the catalog of galaxies used in this work. The ALFALFA team at Cornell is supported by NSF grant AST-0607007 and AST-1107390 and by grants from the Brinson Foundation. M.P. acknowledges the Vice Rector for Research of the University of Innsbruck for the granted scholarship. S.P. acknowledges support from the New Researcher Program (Shinjin grant No. 2019R1C1C1009600) through the National Research Foundation of Korea. S.L. acknowledges the support from the Sejong Science Fellowship Program through the National Research Foundation of Korea (NRF-2021R1C1C2006790). M.B. acknowledges the support from the Polish National Science Centre under the grant 2017/26/D/ST9/00449. 
\end{acknowledgements}

\bibliographystyle{aa}
\bibliography{MATLAS_dwarfsHIv1.bib} 

\onecolumn
\begin{landscape}
\begin{table*}
\tiny
\caption{\label{tab:properties}HI and optical properties of the HI-bearing dwarfs.}
\begin{center}
\begin{tabular}{ccccccccccccccccc}
\toprule
 ID & RA & Dec & v$_{helio}$ & D$_{HI}$ & W$_{50}$ & F$_{HI}$ & SNR & $log(\frac{M_{HI}}{M_{\odot}})$ & M$_g$ & M$_B$ & $(g-r)_0$ & $log(\frac{M_{*}}{M_{\odot}})$ & $\frac{M_{HI}}{L_{B}}$ & $log(\frac{M_{dyn}}{M_{\odot}})$ & Survey & Morph\\
   & [deg] & [deg] & [km/s] & [Mpc] & [km/s] & [Jy-km/s] &  &  & mag & mag & mag &  &  &  &  & \\
(1) & (2) & (3) & (4) & (5) & (6) & (7) & (8) & (9) & (10) & (11) & (12) & (13) & (14) & (15) & (16) & (17)\\
\toprule
MATLAS-23 & 19.7116 & 3.5617 & 4933.0$\pm$8.0 & 71.36$\pm$0.11 & 100 & 1.03$\pm$0.08 & 8.2 & 9.09$\pm$0.06 & -15.08 & -14.76 & 0.29 & 7.88$\pm$0.02 & 18.97 & 10.33 & ALFALFA & dI\\
MATLAS-42 & 20.6260 & 8.7614 & 2422.0$\pm$15.0 & 33.46$\pm$0.21 & 27 & 0.58$\pm$0.05 & 10.7 & 8.18$\pm$0.06 & -15.33 & -14.98 & 0.40 & 8.14$\pm$0.03 & 1.93 & 9.00 & ALFALFA & dI; UDG\\
MATLAS-121 & 26.5425 & 28.8157 & 3764.0$\pm$8.0 & 52.32$\pm$0.11 & 39 & 0.50$\pm$0.04 & 8.4 & 8.51$\pm$0.06 & -14.29 & -13.96 & 0.36 & 7.67$\pm$0.02 & 10.42 & 9.18 & ALFALFA & dE\\
MATLAS-122 & 26.9508 & 22.1426 & 2987.0$\pm$2.0 & 40.63$\pm$0.03 & 36 & 1.40$\pm$0.04 & 21.4 & 8.74$\pm$0.05 & -15.18 & -14.75 & 0.62 & 8.43$\pm$0.01 & 8.45 & 9.82 & ALFALFA & dI\\
MATLAS-124 & 27.2047 & 21.5792 & 3024.0$\pm$6.0 & 42.09$\pm$0.09 & 36 & 0.60$\pm$0.05 & 8.8 & 8.40$\pm$0.06 & -13.76 & -13.47 & 0.21 & 7.24$\pm$0.01 & 12.67 & 9.05 & ALFALFA & dI\\
MATLAS-125 & 27.2088 & 22.1143 & 3039.0$\pm$8.0 & 42.00$\pm$0.12 & 74 & 0.13$\pm$0.03 & 6.2 & 7.70$\pm$0.12 & -15.72 & -15.35 & 0.47 & 8.41$\pm$0.01 & 0.45 & 9.59 & ATLAS$^{3D}$ & dE,N; TTD\\
MATLAS-127 & 27.4368 & 22.3744 & 2947.0$\pm$8.0 & 41.76$\pm$0.12 & 90 & 0.22$\pm$0.06 & 4.8 & 7.94$\pm$0.11 & -15.93 & -15.57 & 0.42 & 8.43$\pm$0.00 & 0.64 & 9.42 & ATLAS$^{3D}$* & dI\\
MATLAS-132 & 27.6409 & 22.0435 & 2347.0$\pm$8.0 & 34.54$\pm$0.12 & 33 & 0.05$\pm$0.03 & 3.5 & 7.17$\pm$0.24 & -10.95 & -10.55 & 0.57 & 6.65$\pm$0.05 & 11.12 & 9.35 & ATLAS$^{3D}$ & dI\\
MATLAS-146 & 29.9250 & 19.0219 & 2372.0$\pm$8.0 & 38.47$\pm$0.12 & 58 & 0.11$\pm$0.04 & 3.6 & 7.69$\pm$0.15 & -15.94 & -15.57 & 0.47 & 8.50$\pm$0.00 & 0.36 & 9.04 & ATLAS$^{3D}$ & dE; TTD\\
MATLAS-154 & 31.7166 & 10.9491 & 1955.0$\pm$8.0 & 28.92$\pm$0.12 & 66 & 0.12$\pm$0.05 & 3.5 & 7.41$\pm$0.18 & -12.96 & -12.6 & 0.42 & 7.24$\pm$0.01 & 2.86 & 8.98 & ATLAS$^{3D}$ & dE\\
... & ... & ... & ... & ... & ... & ... & ... & ... & ... & ... & ... & ... & ... & ... & ...\\
\bottomrule
\end{tabular}
\end{center}
\begin{tablenotes}
    \small
    \item \textbf{Notes.} The columns are as follows: (1) Dwarf ID; (2) Right ascension (J2000); (3) Declination (J2000); (4) Heliocentric velocity; (5) HI distance from Hubble flow; (6) W$_{50}$ line width; (7) HI flux; (8) Signal-to-noise ratio; (9) log10 of the HI mass; (10) g-band absolute magnitude; (11) B-band absolute magnitude; (12) Galactic extinction corrected $g-r$ color; (13) log10 of the stellar mass; (14) HI-mass-to-light ratio; (15) log10 of the dynamical mass; (16) Survey; (17) Morphology. Based on \citet{Poulain2021} for the dEs, dIs, and nucleated (indicated with ",N"), and on \citet{Marleau2021} for the UDGs. ATLAS$^{3D}$* means that the dwarf has also been detected in the ALFALFA survey. This table is available in its entirety at the CDS.
\end{tablenotes}
\end{table*}
\end{landscape}

\begin{appendix}
\section{Dwarfs HI spectra and color cutouts}
\label{AppendixA}
\begin{figure*}[h!]
\centering
\begin{subfigure}{\textwidth}
\includegraphics[page=1,width=\linewidth]{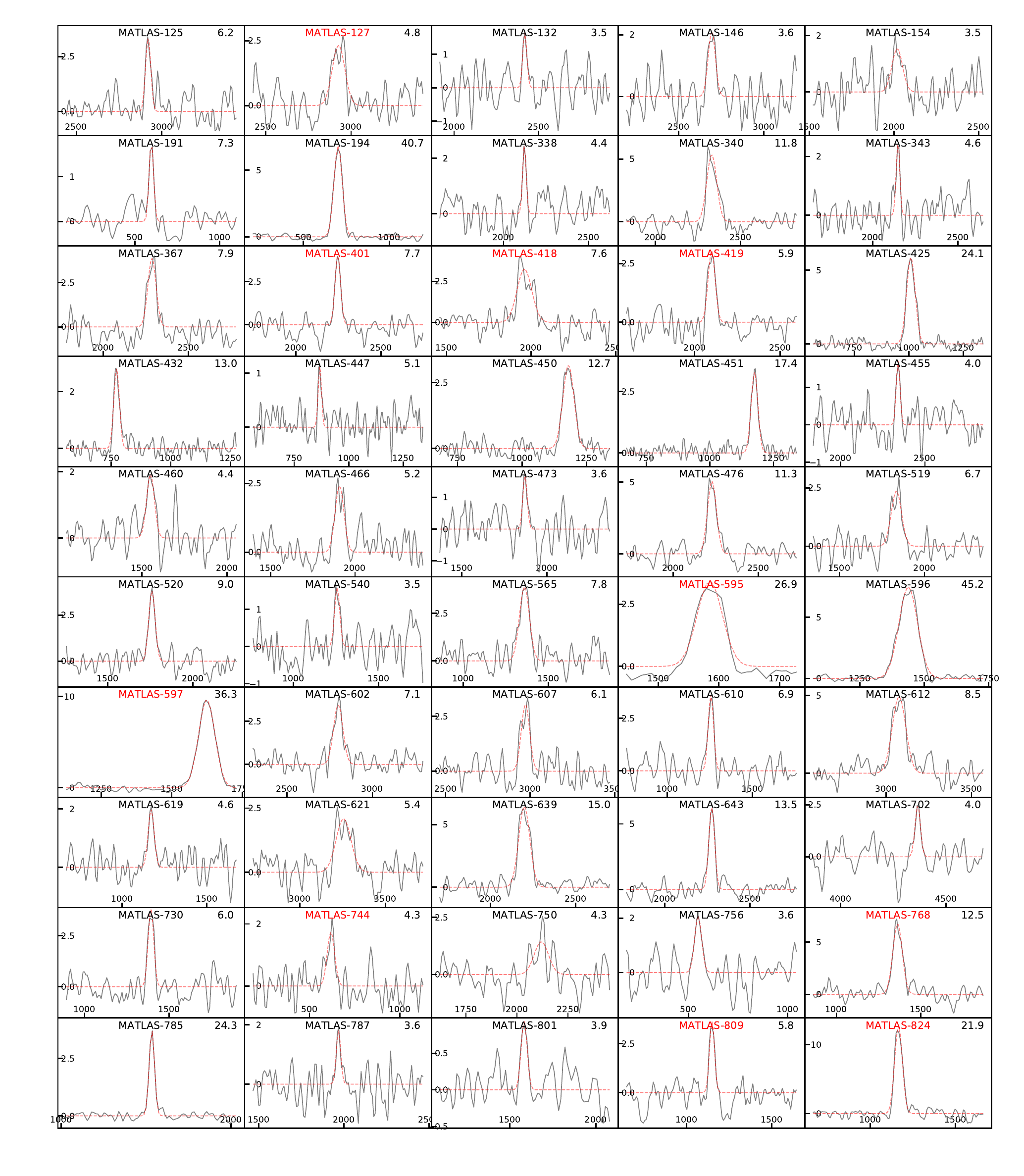}
\end{subfigure}
\caption{HI line spectra of the 94 detected dwarf galaxies in the ATLAS$^{3D}$ HI survey. The flux is represented in mJy. We overplot the Gaussian fit with a red dashed line. Dwarfs detected in both ATLAS$^{3D}$ and ALFALFA surveys have a red name. We indicate the SNR in the top right corner of each spectrum.}
\label{fig:HIspectra}
\end{figure*}

\begin{figure*}
\ContinuedFloat
\centering
\begin{subfigure}{\textwidth}
\includegraphics[page=2,width=\linewidth]{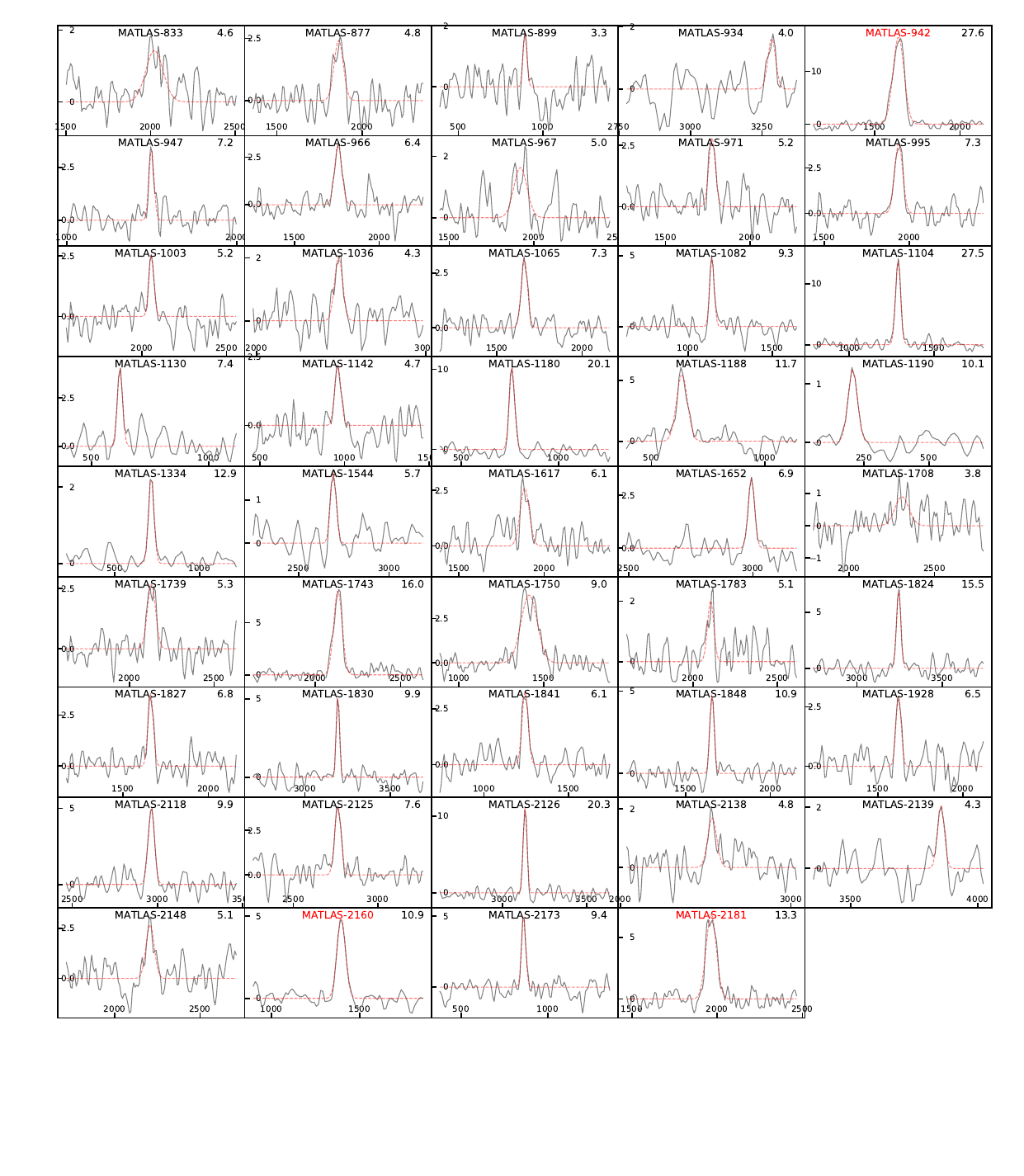}
\end{subfigure}
\caption{Continued.}
\end{figure*}

\begin{figure*}
\centering
\begin{subfigure}{\linewidth}
\includegraphics[width=\linewidth]{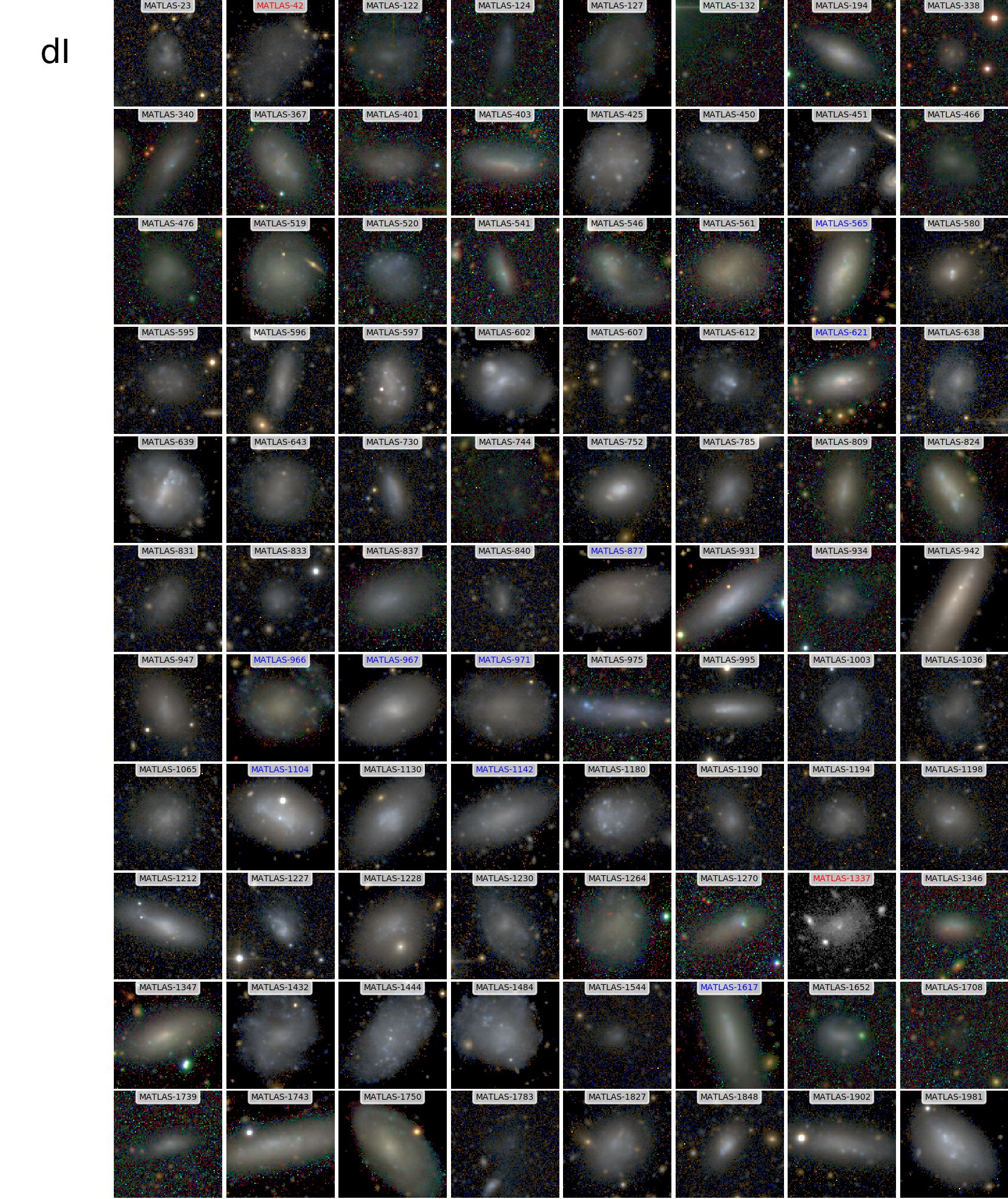}
\end{subfigure}
\caption{Color cutouts of the MATLAS dwarfs with HI detection, divided per morphology (dI, dE). The UDGs have a red name, while the TTD candidates have a blue name. Each cutout is roughly 1\arcmin$\times$1\arcmin with North up and East left, and the RBG images were produced using \textsc{stiff} \citep{Bertin2012}.}
\end{figure*}

\begin{figure*}
\ContinuedFloat
\centering
\begin{subfigure}{\linewidth}
\includegraphics[width=\linewidth]{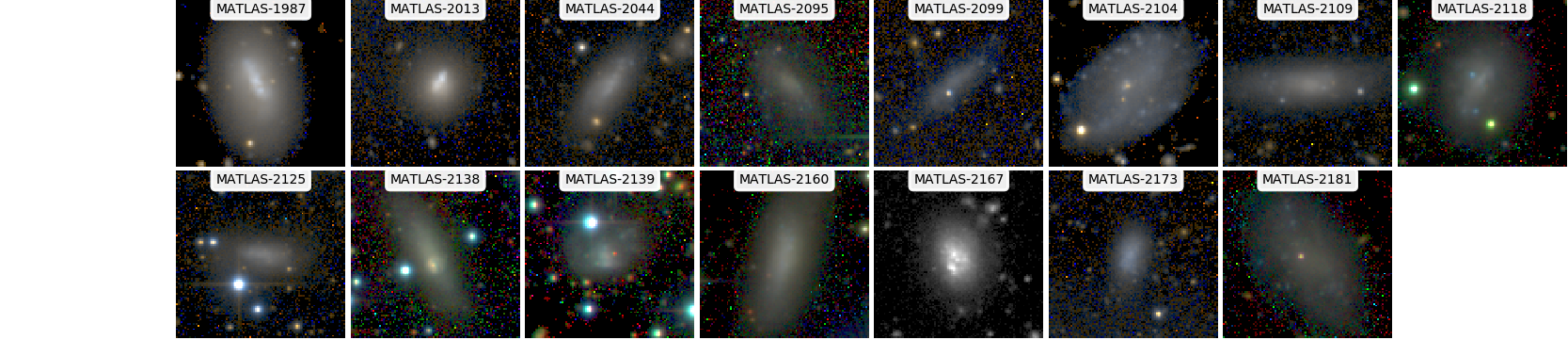}

\vspace{0.5cm}

\includegraphics[width=\linewidth]{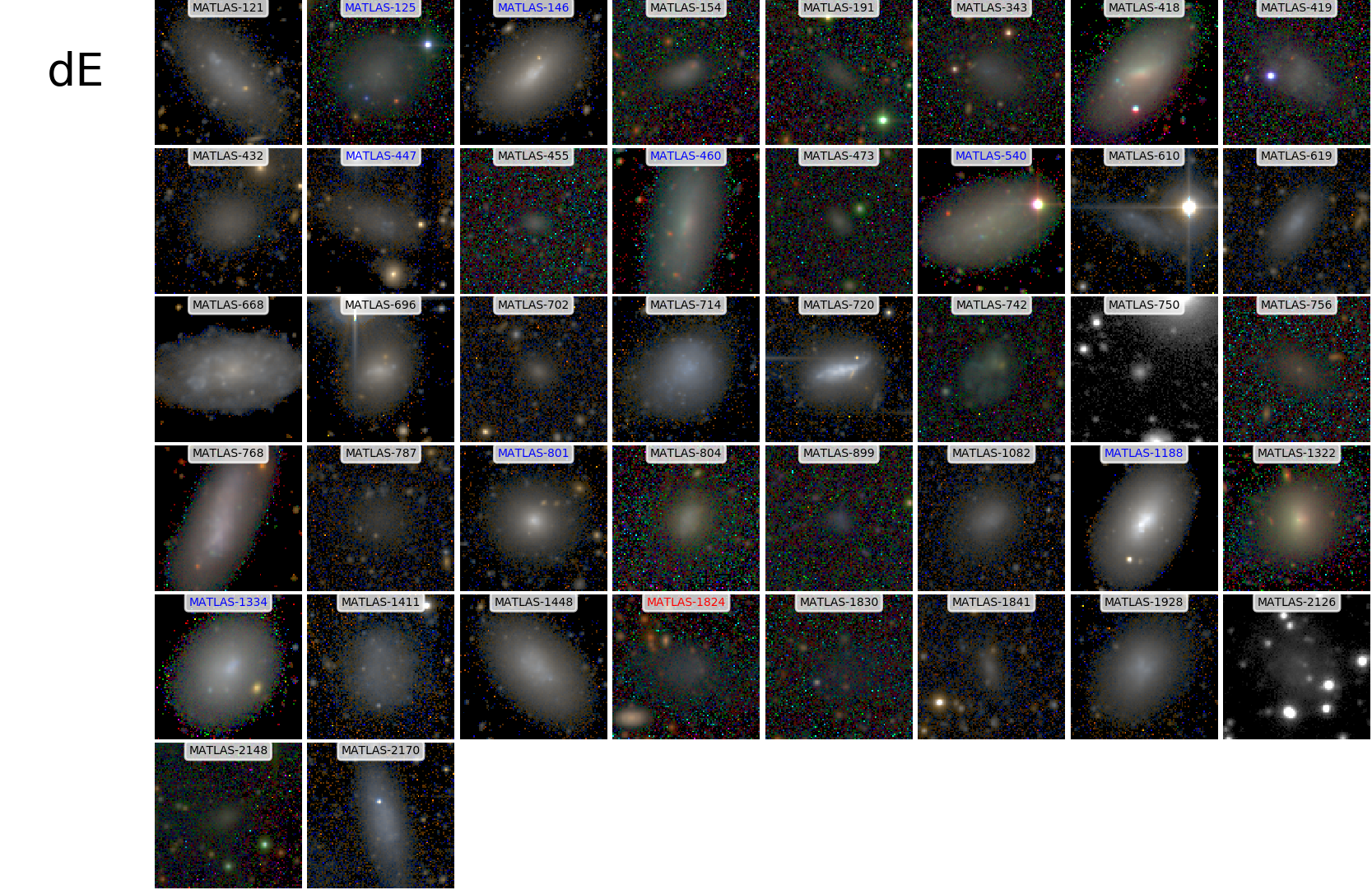}
\end{subfigure}
\caption{Continued}
\label{fig:colorcutouts}
\end{figure*}

\begin{figure}
\centering
\includegraphics[width=0.29\linewidth]{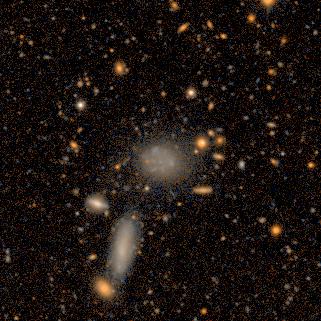}
\includegraphics[width=0.29\linewidth]{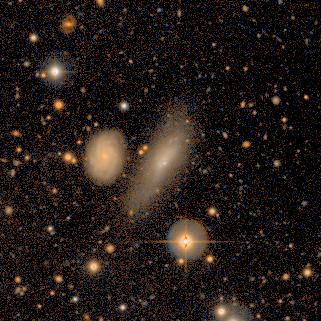}
\includegraphics[width=0.29\linewidth]{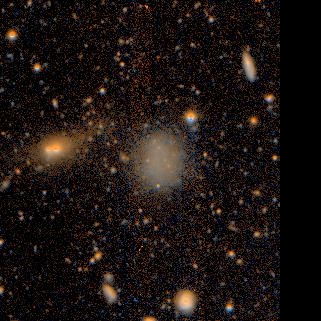}
\includegraphics[width=0.29\linewidth]{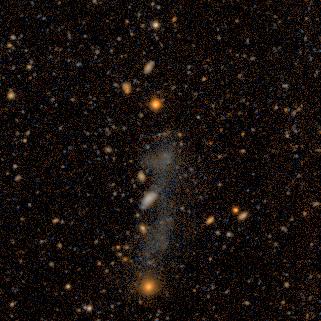}
\includegraphics[width=0.29\linewidth]{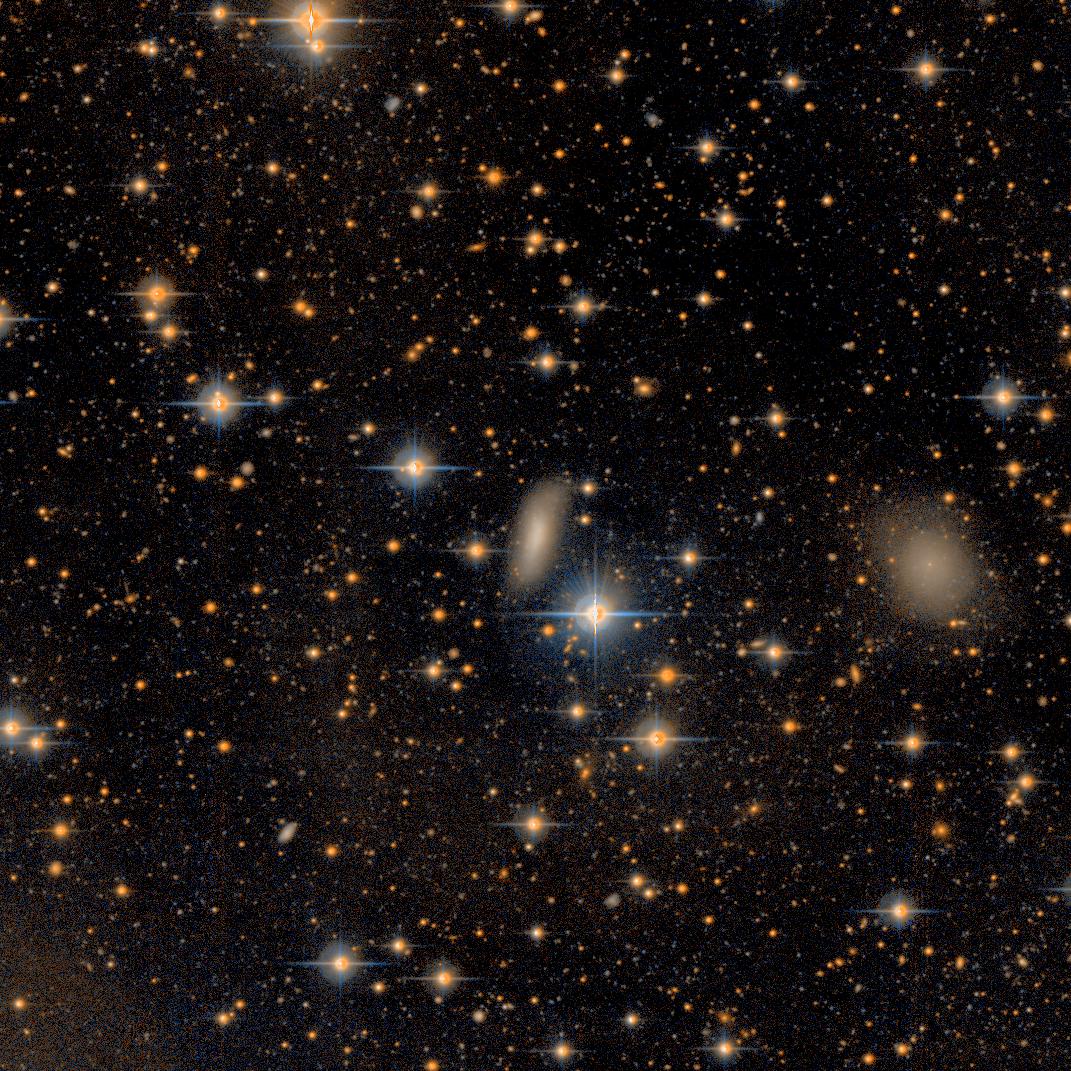}
\includegraphics[width=0.29\linewidth]{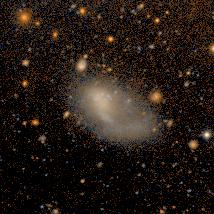}
\includegraphics[width=0.29\linewidth]{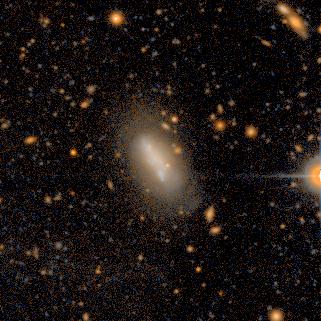}
\includegraphics[width=0.29\linewidth]{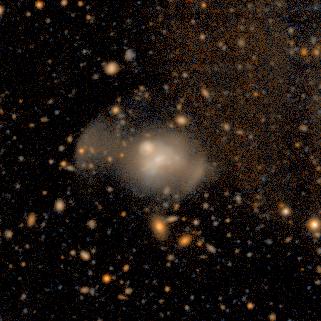}
\includegraphics[width=0.29\linewidth]{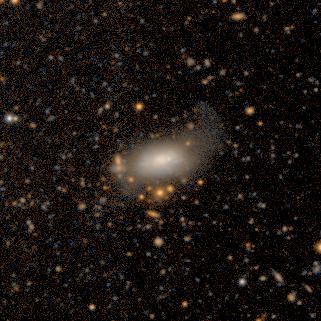}
\includegraphics[width=0.29\linewidth]{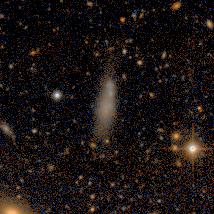}
\includegraphics[width=0.29\linewidth]{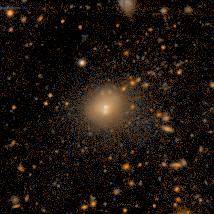}
\includegraphics[width=0.29\linewidth]{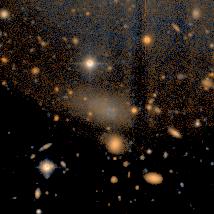}
\caption{Dwarf merger candidates described in Section \ref{section:tidalfeatures}. The two top rows represent the possible pairs, while the dwarfs with shells features are shown in the third row, and the dwarfs with extended isophotes are displayed in the bottom row. From top left to bottom right, the images are centered on the following dwarfs: MATLAS-595, MATLAS-340, MATLAS-1411, MATLAS-1783, MATLAS-2160, MATLAS-546, MATLAS-824, MATLAS-602, MATLAS-621, MATLAS-124, MATLAS-580, MATLAS-447. The four first images are 3\arcmin\ $\times$ 3\arcmin\ , the image in the middle of the second row is 10\arcmin\ $\times$ 10\arcmin\ , and the remaining images are 2\arcmin\ $\times$ 2\arcmin\ with North up and East left. RGB images were produced with the help of the \textsc{Astropy} package, based on the method from \citet{Lupton2004}.}
\label{fig:mergers}
\end{figure}
\end{appendix}

\end{document}